\documentclass[lettersize,journal]{IEEEtran}
\usepackage{textcomp}
\usepackage{stfloats}
\usepackage{url}
\usepackage{verbatim}
\usepackage{graphicx}
\usepackage{cite}
\usepackage{hyperref}
\hypersetup{hidelinks,
	colorlinks=true,
	allcolors=black,
	pdfstartview=Fit,
	breaklinks=true}
\usepackage{indentfirst}
\usepackage{amsmath,amssymb,amsfonts}
\usepackage{mathabx}
\usepackage[version=4]{mhchem}
\usepackage{mathrsfs}
\usepackage{xcolor}
\graphicspath{{figures/}}
\usepackage{graphbox}
\usepackage{float} 
\usepackage{caption2} 
\usepackage{booktabs}
\usepackage{subfigure}
\usepackage{amsthm}
\usepackage{lettrine}
\usepackage{ragged2e}
\usepackage{makecell}

\newtheorem{definition}{Definition}
\usepackage{pifont}
\usepackage[ruled, vlined, linesnumbered]{algorithm2e}

\usepackage{float}  
\usepackage{lipsum}
\usepackage{tabularx}      
\usepackage{array}         
\usepackage{booktabs}      
\newcolumntype{L}{>{\raggedright\arraybackslash}X}

\usepackage{enumitem}

\makeatletter

\newboolean{include-notes}
\setboolean{include-notes}{true}
\newcommand{\nb}[3]{\ifthenelse{\boolean{include-notes}}{{\colorbox{#2}{\bfseries\sffamily\scriptsize\textcolor{white}{#1}}}{\ \textcolor{#2}{\sf\small\textit{#3}}}}{}}

\makeatother
\usepackage{makecell}

\begin{document}


\title{Hybrid Orchestration of Edge AI and Microservices via Graph-based Self-Imitation Learning}
\author{Yang Chen, Zheng Jin, Zhuolin Yang, Pan Lai, Xiao Zhang, Menglan Hu and Haiyan Yin

\thanks{Yang Chen, Zheng Jin, Zhuolin Yang, Pan Lai, and Xiao Zhang are with School of Computer Science, South-Central Minzu University, Wuhan, China, 430074, and Key Laboratory of Cyber-Physical Fusion Intelligent Computing, State Ethnic Affairs Commission. E-mails:  \{2023110231, 2024120403, 2024120445\}@scuec.edu.cn, plai1@ntu.edu.sg, xiao.zhang@my.cityu.edu.hk.}

\thanks{Menglan Hu is with Hubei Key Laboratory of Smart Internet Technology, School of Electronic Information and Communication, Huazhong University of Science and Technology, Wuhan, China, 430074. E-mails: humenglan@hust.edu.cn.}

\thanks{Haiyan Yin is with Centre for Frontier AI Research, Agency for Science, Technology and Research (A*STAR), Singapore. E-mails: yin\_haiyan@cfar.a-star.edu.sg.}

}





\maketitle
\begin{abstract}
\justifying
Modern edge AI applications increasingly rely on microservice architectures that integrate both AI services and conventional microservices into complex request chains with stringent latency requirements. Effectively orchestrating these heterogeneous services is crucial for ensuring low-latency performance, yet remains challenging due to their diverse resource demands and strong operational interdependencies under resource-constrained edge environments. 
In particular, frequent interactions between services tightly couple deployment and routing decisions, yet existing approaches optimize them in isolation, leading to fundamentally inadequate system performance.
In this paper, we propose \textbf{SIL-GPO}, a reinforcement learning framework that optimizes hybrid orchestration for edge AI microservice systems. SIL-GPO formulates the orchestration problem as a sequential decision-making task and leverages graph attention networks to encode service topologies and routing dependencies within the agent state representation. 
Moreover, SIL-GPO integrates a self-imitation learning strategy into proximal policy optimization, enabling the agent to prioritize and reuse high-reward trajectories. This guides policy updates towards globally promising solutions that standard RL often fails to discover under sparse rewards and large combinatorial action spaces.  
We conduct extensive experiments on trace-driven edge AI workloads, demonstrating that SIL-GPO significantly reduces end-to-end service latency and enhances resource utilization compared to state-of-the-art heuristic, metaheuristic, and deep RL baselines. 
Our framework offers a unified and scalable solution for efficient orchestration of AI services and microservices in the edge, paving the way for low-latency, high-performance edge AI deployments.
\end{abstract}
\section{Introduction}

\IEEEPARstart{R}{ecently}, the rapid advancement of AI has spurred its widespread adoption across multiple industries \cite{introduction_b1}. Currently, major cloud service providers, including Google, AWS, and IBM, offer comprehensive suites of cloud-based AI services that span the full spectrum of AI research and industrial deployment \cite{introduction_b2}. However, these AI applications, such as intelligent driving assistance systems and industrial automation systems, impose stringent requirements on service response time. The inherent high and variable transmission delay associated with cloud servers renders these applications poorly suited for cloud deployment. Edge computing addresses this limitation by enabling service provisioning at the network edge, thereby delivering low-delay and stable services. Furthermore, the emergence of Model as a Service (MaaS) has facilitated the practical deployment of AI services at the edge \cite{introduction_b3}.
\begin{figure}[t]
    \centering
    \includegraphics[width=88.9mm]{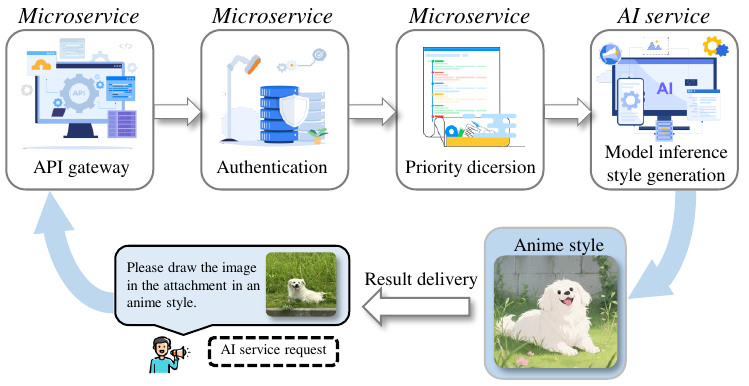}
    \vspace{-0.5cm}
    \caption{An example of image-generating intelligent application.}
    \label{video}
    \vspace{-0.7cm}
\end{figure}

Leveraging the MaaS paradigm, AI service providers deploy large-scale models on edge servers to achieve AI as a service (AIaaS). Currently, through microservice architecture, an AI application consists of multiple microservice components and AI services, which are deployed in containers to achieve rapid instantiation \cite{introduction_b4}. Service requests generated by AI applications encompass not only AI-specific services but also conventional microservices implementing fundamental functionalities, such as API gateway management and control validation \cite{ex}. However, existing research predominantly equates services in AI applications solely with AI services \cite{introduction_b5} \cite{introduction_b6}, focusing exclusively on optimizing AI service deployment while neglecting non-AI microservices. Consequently, investigating the hybrid orchestration of AI services and microservices is imperative to achieve comprehensive deployment of AI application services and enhance the user-perceived quality of service. Nevertheless, significant challenges arise in this hybrid orchestration due to operational dependencies (e.g., data flow coupling) and fundamental incompatibilities (e.g., resource isolation mechanisms) between AI services and microservices.

The tight coupling between AI services and microservices, coupled with the inherent resource constraints of edge servers, significantly increases the complexity of hybrid orchestration. Typically, AI applications require the execution of one or more prerequisite microservice components before the AI service itself can commence \cite{introduction_b9}. For instance, as illustrated in Fig. \ref{video}, prior to accessing AI services for image generation, intelligent image generation applications must invoke the gateway authentication microservice to authorize access request connections. Therefore, processing user service requests requires multiple network calls, inevitably introducing significant delays in cross-service communication. Given that communication overhead between internal components of a server is negligible, service aggregation, i.e., the deployment of interdependent services on the same server, constitutes an effective strategy for mitigating these communication delays. However, resource-constrained edge servers typically lack of the capacity to support large-scale service aggregation. Furthermore, resource isolation between services inhibits resource sharing, inevitably resulting in resource competition among services. Specifically, to ensure service quality, microservices commonly employ multiple lightweight instances \cite{introduction_b7}, whereas AI services typically require only a single instance for efficient inference but demand significant computational resources \cite{introduction_b8}. This fundamental difference in deployment paradigms, combined with the high dependence of AI services on scarce GPU resources, underscores the critical need for refined, collaborative deployment strategies. Therefore, optimizing the hybrid orchestration of AI services and microservices to enhance service aggregation efficiency and minimize resource contention is paramount to improving overall service quality.

Furthermore, efficient hybrid service orchestration relies on joint optimization of service deployment and request routing. Crucially, the efficiency of request routing is fundamentally dependent on the underlying deployment topology of service instances and is highly sensitive to their fine-grained placement strategies \cite{introduction_b11}. Currently, the performance of the service deployment itself depends on the scheduling and routing decisions made between service instances \cite{introduction_b12}. Joint optimization of service deployment and request routing effectively reduces both processing and communication latencies, thereby significantly minimizing the overall latency for user service requests. However, current research often focuses exclusively on optimizing the deployment of the AI service \cite{introduction_b13} \cite{introduction_b14} \cite{introduction_b15}, neglecting the critical aspect of request routing optimization. Thus, achieving efficient hybrid orchestration of AI services and microservices requires concurrent optimization of both deployment placement and request routing strategies.

To address these challenges, we propose a reinforcement learning algorithm based on Graph Attention Networks (GATs) for the hybrid orchestration of AI services and microservices, with the objective of optimizing response delay. Specifically, our contributions are summarized as follows:

\begin{itemize}
    \item We develop a \textbf{fine-grained multi-instance hybrid orchestration model} based on the Open Jackson queuing network. This model captures both service processing delays and inter-service communication latencies, thereby enabling end-to-end delay optimization for AI-microservice deployment. Subsequently, we formalize the \textbf{hybrid orchestration of AI services and microservices} as a mixed integer nonlinear programming problem, and model it as a sequential decision-making task.

    \item We propose the Self-Imitation Learning-enhanced Graph Policy Optimization Algorithm (SIL-GPO) to solve it. Our method integrates \textbf{GATs} to encode service dependencies with \textbf{self-imitation learning} to reuse high-reward trajectories, enhancing state representations, accelerating convergence, and improving exploration efficiency in large combinatorial action spaces.
    
    \item We conduct extensive trace-driven experiments demonstrating that our approach significantly reduces service delay and improves resource utilization compared to state-of-the-art heuristic, metaheuristic, and deep reinforcement learning baselines. Specifically, in comparison with the optimal baseline algorithm, SIL-GPO demonstrated a reduction in total response delay of 15.19\%.
\end{itemize}

The remainder of this paper is organized as follows. Section \ref{section2} reviews related work on service orchestration. Section \ref{section3} presents the edge AI microservice system model and formulates the hybrid services orchestration optimization problem. Section \ref{section4} details SIL-GPO, our proposed graph-based reinforcement learning algorithm for hybrid orchestration. Section \ref{section5} reports the simulation results and performance analysis. Finally, Section \ref{section6} concludes the paper. 
\section{Related work}\label{section2}
\subsection{Service Deployment Strategies}
The hybrid orchestration of AI services and microservices aims to address the challenge of efficient co-deployment. However, existing research has primarily focused on the efficient deployment of either AI services or microservices in isolation, with limited consideration given to their joint orchestration.
\subsubsection{Deployment of Conventional Microservices}
Conventional microservice deployment research predominantly addresses the horizontal or vertical scaling of microservice instances. Lv \emph{et al.} \cite{related_microservice_deployment_b1} formulated a multi-objective deployment problem for edge computing and employed a Reward Sharing Deep Q-Learning (RSDQL) algorithm, building upon a heuristic elastic scaling approach to manage dynamic request loads. Zeng \emph{et al.} \cite{related_microservice_deployment_b2} introduced SafeDRL, a reinforcement learning algorithm designed to efficiently manage dynamically arriving user requests while meeting reliability and latency constraints. Chen \emph{et al.} \cite{related_microservice_deployment_b3} tackled service deployment in heterogeneous, dynamic edge-cloud environments using a multi-buffer depth-determined policy gradient algorithm. Fu \emph{et al.} \cite{related_microservice_deployment_b4} focused on business-oriented deployment, presenting a QoS-based deployment model and utilizing an immune genetic algorithm for its solution.

\subsubsection{Deployment of AI Service}
Research on deploying AI services primarily focuses on addressing challenges associated with large-scale, resource-intensive inference models. Wu \emph{et al.} \cite{introduction_b5} decoupled AI services into multiple DNN models and proposed the TDDP algorithm to generate cost-effective schemes for orchestrating multiple DNNs and allocating load. He \emph{et al.} \cite{related_AI_service_deployment_b1} proposed a heuristic solution based on the Dink Beetle algorithm for the joint optimization of model selection, service placement, user access, and splicing coefficients. Chen \emph{et al.} \cite{related_AI_service_deployment_b2} proposed a hybrid active-passive automatic scaling strategy for PRHAS, which optimizes scaling strategies by predicting future traffic and startup times for AI microservices. Niu \emph{et al.} \cite{related_AI_service_deployment_b3} proposed ChainNet to assess the reliability of alternative deployments and to guide a loss-aware search process for finding the optimal edge AI deployment scheme. Li \emph{et al.} \cite{related_AI_service_deployment_b4} achieved the lowest-cost automated real-time cloud deployment for online DNN inference under latency constraints by jointly utilizing Bayesian optimization and deep reinforcement learning, and improved deployment efficiency by employing a detection-based notification block reuse mechanism and a tensor algebra super optimizer.


Despite these valuable contributions addressing the deployment challenges of individual AI services or microservices effectively, limited attention has been paid to the interdependencies between them within AI applications. To bridge this gap, this paper introduces a fine-grained hybrid multi-instance queuing model designed to facilitate the efficient hybrid orchestration of AI services and microservices.

\subsection{Request Routing Approaches}
Resource-constrained edge servers are unable to independently process complete service requests, necessitating collaborative routing across multiple edge servers for request completion. An efficient routing scheduling strategy can significantly reduce communication delay overhead induced by inter-server coordination. Consequently, designing efficient request routing algorithms has become a focal point for researchers in edge computing.

Harahap \emph{et al.} \cite{related_routing_b1} proposed a router-based request routing model for Content Delivery Networks (CDNs), embedding redirection logic within service-oriented routers to minimize end-to-end latency. Wang \emph{et al.} \cite{related_routing_b2} introduced a two-timescale optimization framework for multi-edge clusters, dynamically balancing latency and scaling costs through joint cluster scaling and request routing. Zeng \emph{et al.} \cite{related_routing_b3} designed an adaptive Savitzky-Golay filter–LSTM–attention prediction algorithm (ASLA) coupled with a differentiated routing mechanism, leveraging service-network coordination to implement differentiated request routing (DCSN-DR) in video CDNs. Zhao \emph{et al.} \cite{related_routing_b4} developed a distributed redundant layout framework (SAA-RP), scheduling service candidates to multiple sites to accelerate request response.

While these studies advance request routing optimization under diverse network and service conditions, they have not sufficiently addressed the tight coupling between request routing and service deployment. In contrast, our method addresses the joint optimization of service deployment and request routing, further proposing a fine-grained hybrid multi-instance queuing model to holistically analyze end-to-end request processing.
\subsection{Joint Optimization of Deployment and Routing}
With the advancement of research on network service quality, joint optimization of service deployment and request routing has garnered significant attention in academia. The core challenge stems from the tight coupling between service deployment and routing decisions, which complicates their decomposition into simpler, independent subproblems. To address this challenge, several researchers have proposed dedicated solutions.

Somesula \emph{et al.} \cite{related_joint_b1} addressed collaborative service placement and routing in resource-constrained mobile edge networks using integer linear programming combined with greedy rounding, randomized rounding, and a circular convex set-based heuristic algorithm. Chen \emph{et al.} \cite{related_joint_b2} developed a dynamic service migration framework for MEC networks, leveraging Lyapunov optimization and randomized rounding techniques to enhance service deployment and routing decisions. Hu \emph{et al.} \cite{related_joint_b3} formulated the problem as a mixed-integer nonlinear program and proposed a two-stage heuristic algorithm based on resource partitioning and partition mapping, aimed at minimizing total network overhead and latency. Xu \emph{et al.} \cite{related_joint_b4} introduced a genetic algorithm with local search for microservice instance deployment and a probabilistic forwarding strategy for request routing, optimizing microservice system performance. Peng \emph{et al.} \cite{related_joint_b5} designed an approximate rounding-based algorithm to jointly optimize service deployment and request routing, targeting service response delay minimization.

While these studies provide valuable insights into the joint optimization of microservice deployment and routing, they primarily employ a two-stage decoupled approach. More critically, existing works largely overlook the unique challenges of AI services, such as complex model architectures and specialized GPU resource requirements. In contrast, this work addresses the hybrid deployment of conventional microservices and AI microservices and jointly optimizes deployment and routing strategies within a unified framework.
\section{AI Service and Microservice Hybrid Orchestration Problem Formulation}\label{section3}
\begin{table}[htbp]\scriptsize 
    \renewcommand{\arraystretch}{1.5}
    \caption{Notation and terminology}
    \label{Notations}
    \centering
    \begin{tabularx}{0.47\textwidth}{l L}
        \hline\hline
        \textbf{Notations} & \textbf{Definitions} \\ 
        \hline
        $\mathcal{G}(N,E)$ & The edge AI service network   \\ \hline
        $R_n^{res}$ & The size of heterogeneous resources owned by server $n$ \\ 
        \hline
        $B_{ij}$ & The bandwidth between server $n_i$ and server $n_j$   \\ 
        \hline 
        $s, S$ & The service instance and the set of service instances \\ 
        \hline
        $sc, SC$ & The service request and The set of service requests \\ 
        \hline
        $r_{s}^{res}$ & The size of heterogeneous resources consumed by service $s$ \\ 
        \hline
        $N_{n}^{s}$ & The number of service instances $s$ on server $n$ \\
        \hline
        ${{p}({sc})}_{n_i,s_i}^{n,s}$ & Probability that the request $sc$ is routed from service $s$ on server $n$ to service $s_i$ on server $n_i$ \\
        \hline
        $\lambda _{n}^{s}$ & The arrival rate of service $s$ on server $n$ \\ 
        \hline
        ${RP(sc)}$ & The set of routing paths for service request $sc$\\ 
        \hline
        ${P(sc)_{rp(sc)}}$ & Probability of routing path for service request $sc$\\ 
        \hline
        $D_{sc}$ & The packet size of service request $sc$ \\ \hline
        $\mu_s$ &  The processing rate of service $s$ \\ 
        \hline
        $D_s$ & The size of the result data for service $s$ \\
        \hline
        $D_{sc}^{return}$ & The size of the result data for service request $sc$ \\ 
        \hline
        $T_{sc}^{t}$ & Transmission delay of service request $sc$\\ 
        \hline
        $T_{n,s}^{q,p}$ & Queuing and processing delay of service $s$ on server $n$\\ 
        \hline
        $T_{n,n'}^{c}$ & Communication delay between server $n$ and $n'$\\ 
        \hline
        $T_{sc}^{r}$ & Return delay of service request $sc$ \\ \hline
        $\mathbb{E}[T_{sc}]$ & The average service request response delay \\ \hline\hline
        \end{tabularx}
\end{table}
\begin{figure*}[htbp]
    \centering
    \includegraphics[width=182mm]{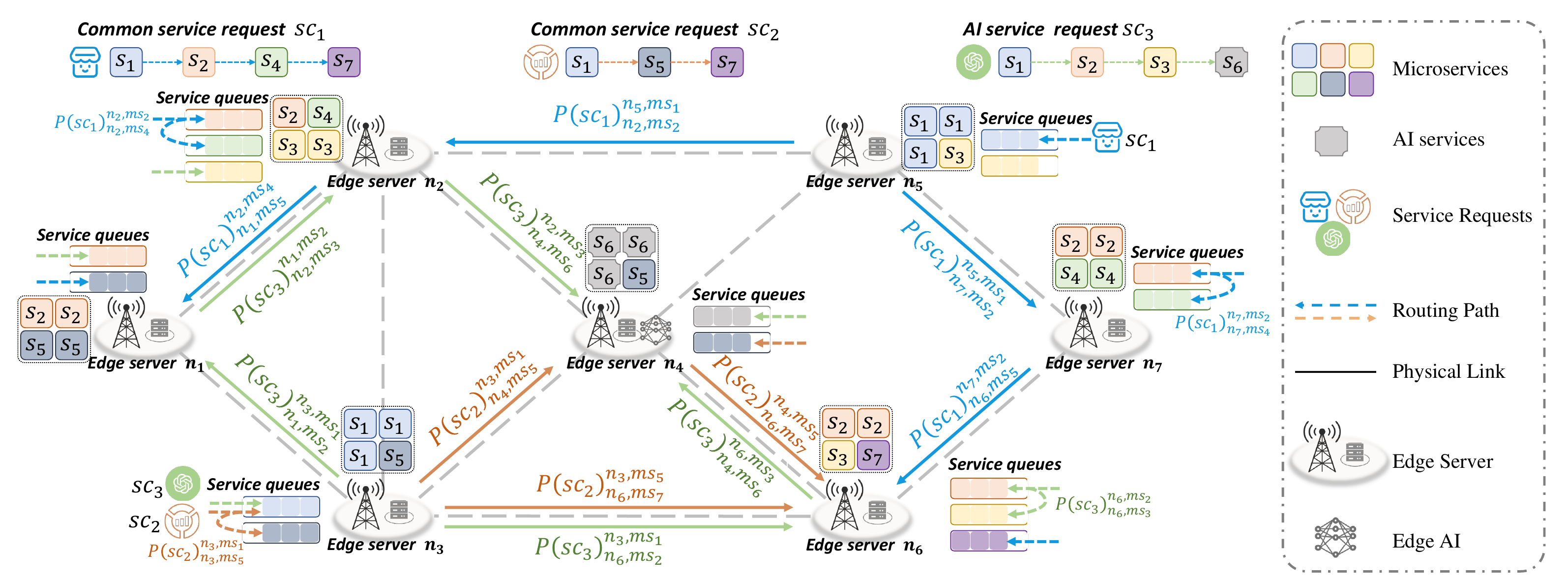}
    \caption{Diagram of hybrid deployment and request routing for AI service and microservice.}
    \label{network}
    \vspace{-0.3cm}
\end{figure*}
\subsection{Edge Service Network System Models}
The proposed AI service network architecture comprises a set of heterogeneous edge servers, modeled as an undirected graph $\mathcal{G}(N, E)$. Here, $N$ represents the set of heterogeneous edge servers and $E$ denotes the set of communication links connecting them. Based on the type of computational resources they provide, the edge servers $N$ are categorized into two disjoint subsets: Universal-Computing-Storage (UCS) servers $N_{UCS}$ and Hybrid-Accelerated-Computing (HAC) servers $N_{HAC}$, satisfying $N = N_{UCS} \bigcup N_{HAC}$. Each Universal-Computing-Storage server $n_{ucs} \in N_{UCS}$ possesses CPU and memory resources of fixed capacity. In contrast, each Hybrid-Accelerated-Computing server $n_{hac} \in N_{HAC}$ offers augmented CPU and memory resources at fixed capacities, along with supplementary GPU resources. Fig. \ref{network} illustrates the network topology, wherein edge server $n_4$ is a centrally located Hybrid-Accelerated-Computing server, while all other nodes are Universal-Computing-Storage servers. For brevity in subsequent discussion, the terms ``Universal-Computing-Storage server" and ``Hybrid-Accelerated-Computing server" are collectively referred to as ``server".

The resource capacity of server $n \in N$ is defined by $R_n^{res} = (R_n^{cpu}, R_n^{gpu}, R_n^{mem})$, where: $R_n^{cpu}$ denotes the available CPU resource units on server $n$, $R_n^{gpu}$ denotes the available GPU resource units on server $n$, and $R_n^{mem}$ denotes the available memory resource units on server $n$. For a Universal-Computing-Storage server ($n \in N_{UCS}$), $R_n^{gpu} = 0$. Conversely, for a Hybrid-Accelerated-Computing server ($n \in N_{HAC}$), $R_n^{gpu} > 0$. Furthermore, each inter-server link possesses dedicated bandwidth resources. The bandwidth between server $n_i$ and server $n_j$ is denoted by $B_{n_i,n_j}$.

\subsection{Multi-Service Hybrid Service Request Definitions}
The service ecosystem within the network comprises two distinct types: AI services and microservices. Due to computational resource limitations on edge servers, AI services must be hosted exclusively on HAC servers, as their execution necessitates GPU hardware. Conversely, microservices can be deployed on any edge server within the network infrastructure. Fig. \ref{network} exemplifies this constraint, where AI service $s_6$ is solely deployable on edge server $n_4$, while microservices are deployed across all server types. Next, we provide the mathematical definitions for microservices and AI services respectively.

\begin{definition}
    (Microservices instances) The set of microservices is denoted by $S^{MS} = \{s_{1}^{ms}, s_{2}^{ms}, \dots\}$, with $\left|S^{MS}\right|$ denotes the number of types of microservices in the network. The resource requirements of microservice $s_{i}^{ms}$ are represented as a triplet $r_{s_{i}^{ms}}^{res}=(r_{s_{i}^{ms}}^{cpu}, r_{s_{i}^{ms}}^{gpu}, r_{s_{i}^{ms}}^{mem})$, denoting the requirements for CPU, GPU, and storage resources, respectively. It is worth noting that $r_{s_{i}^{ms}}^{gpu}=0$ holds true at all times. Additionally, the processing rate of microservice $s_{i}^{ms}$ is $\mu_{i}^{ms}$, and the size of the resulting data is $D_{i}^{ms}$.
\end{definition}

\begin{definition}
    (AI services instances) The set of AI services is denoted by $S^{AI} = \{s_{1}^{ai}, s_{2}^{ai}, \dots\}$, with $\left|S^{AI}\right|$ denotes the number of types of microservices in the network. The resource requirements of microservice $s_{i}^{ai}$ are represented as a triplet $r_{s_{i}^{ai}}^{res}=(r_{s_{i}^{ai}}^{cpu}, r_{s_{i}^{ai}}^{gpu}, r_{s_{i}^{ai}}^{mem})$, denoting the requirements for CPU, GPU, and storage resources, respectively. Additionally, the processing rate of microservice $s_{i}^{ai}$ is $\mu_{i}^{ai}$, and the size of the resulting data is $D_{i}^{ai}$. Due to the unique internal processing mechanisms of AI services, their storage resource requirements $r_{s_{i}^{ai}}^{mem}$ and processing rates $\mu_{i}^{ai}$ will be detailed separately in subsection \ref{AI service inference}.
\end{definition}


Services do not arrive as individual units but rather as bundled service requests. Each service request comprises multiple AI services and microservices. The mathematical definition of a service request is as follows:
\begin{definition}
    (Service requests) The set of service requests is denoted by $SC = \{sc_1,sc_2,\dots\}$. For each service request $sc_i$, its service chain is defined as $ch(sc_i)=\{s_1,s_2, \dots\}$. Additionally, the transmission size of service request $sc$ is $D_{sc}$, and the expected result size is $D_{sc}^{return}$.
\end{definition}

Based on their functionality, service requests are categorized as AI service requests or common service requests. If the service request $sc$ is an AI service request, then the request chain $ch(sc)$ of service request $sc$ must satisfy the following conditions: $\exists s_{j} \in ch(sc),s_{j}\in S^{AI}$; Conversely, if the service request $sc$ is a regular service request, then the request chain a of service request $ch(sc)$ must satisfy the following conditions: $\forall s_{j} \in ch(sc),s_{j}\in S^{MS}$.
As depicted in Fig. \ref{network}, three service requests exist in the network, of which $sc_1$ and $sc_2$ are common service requests, and $sc_3$ is an AI service request, where the request chain of $sc_3$ is $ch(sc_3)=\{s_1,s_2,s_3,s_6\}$ which includes the AI service $s_6$. 
\subsection{Analysis of AI Service Inference Efficiency}
\label{AI service inference}
Edge AI services rely heavily on the underlying LLM for their task-processing capabilities. LLM inference is decomposed into two stages: context prefill and autoregressive decoding. In the context prefill stage, the model performs full-context encoding to construct the key–value cache(KV cache) for subsequent autoregressive decoding. During the autoregressive decoding stage, each output token is generated sequentially based on the previously generated tokens and the cached key–value states. This paper will analyze the computational load during edge AI service inference using the open-source LLAMA3 model as an example. 

The LLAMA3 model employs a group query mechanism in its Transformer layers and features FFN layers with a SwiGLU architecture. Assume that the input sequence length of the request received by the AI service $s^{ai}$ is $s_{in}$, and the output sequence length is $s_{out}$. The hidden layer dimension of each Transformer layer in the model architecture is $H$, and the total number of attention heads is $d$. Therefore, the dimension of a single attention head is $h$, where $h = H/d$. Additionally, the number of groups in a Transformer layer is $g$, that is, the number of KV heads is $g$. It is worth noting that here we focus primarily on the computational load generated by all Transformer layers within the model, disregarding the minor computational overhead associated with vector embeddings and other transformations.

\textbf{Prefill stage:} The input representation of an attention head in the Transformer layer is denoted as $X\in \mathbb{R}^{s_{in}\times H}$. Then, The query projection matrix, key projection matrix, and value projection matrix of the Transformer layer are denoted as $W_Q\in \mathbb{R}^{H\times H}$, $W_K\in \mathbb{R}^{H\times h_g}$, and $W_V\in \mathbb{R}^{H\times h_g}$, respectively, where $h_g=gh$. Subsequently, the query, key, and value calculations are as follows:
\begin{equation}
    Q=X\cdot W_Q, \quad K=X\cdot W_K, \quad V=X\cdot W_V
\end{equation}

Therefore, the computational overhead of the query, key, and value can be expressed as $F_{qkv}^{prefill}$:
\begin{equation}
    F_{qkv}^{prefill} = 2s_{in}H^2+4s_{in}Hh_g = 2s_{in}H^2+4s_{in}Hgh
\end{equation}

Subsequently, Q will be divided into multiple $Q^h$ with a total of $d$ of them. $K$ and $V$ will be divided into $g$ groups. And every $d_g$ quantity of $Q^h$ shares the same group of $K^h$ and $V^h$, where $d_g = d/g$. Additionally, for each $Q^h$ and $K^h$, the positional encoding will undergo RoPE (Relative Position Embedding) rotation, and then the output of each attention head will be calculated. Assuming that the $i$-th attention head corresponds to the $j$-th group of KV. Then, $j =\lceil i/d_g\rceil$. Therefore, the output of the $i$-th attention head is:
\begin{equation}
    A_i^h=softmax\left(\frac{Q_i^h\cdot \left(K_j^h\right)^\mathsf{T}}{\sqrt{h}}\right)\cdot V_j^h
\end{equation}

Subsequently, the outputs of all the attention heads are integrated through the projection matrix $W_O\in \mathbb{R}^{H\times H}$:
\begin{equation}
    X_{out} = concat\left(A_1^h,A_2^h,\dots,A_d^h\right)\cdot W_O+X
\end{equation}

Therefore, the computational cost of all the attention head output data can be expressed as $F_{attn}^{prefill}$ :
\begin{equation}
    F_{attn}^{prefill} = 4{s_{in}}^2{h}d+2s_{in}H^2 = 4{s_{in}}^2H+2s_{in}H^2
\end{equation}

Finally, through the SwiGLU gating mechanism and the dimensionality increase and decrease processing of the FFN layer, the output of the current transformer layer is obtained:
\begin{equation}
    X_{gate}=\text{SiLU}\left(X_{out}\cdot W_1\right)\otimes\left(X_{out}\cdot W_3\right)
\end{equation}
\begin{equation}
    X_{final} = X_{gate}\cdot W_2+X_{out}
\end{equation}
where $W_1\in \mathbb{R}^{H\times H}$ and $W_3\in \mathbb{R}^{H\times mH}$ is the projection matrix increasing in dimension and $w_2 \in \mathbb{R}^{mH\times H}$ is the projection matrix reducing in dimension.

Therefore, the computational load in the feed forward network is expressed as $F_{feed}^{prefill}$:
\begin{equation}
    F_{feed}^{prefill} = 6m{s_{in}}H^2
\end{equation}

In summary, in the prefill stage, the inference computational cost of a single Transformer layer is expressed as:
\begin{equation}
\begin{aligned}
    F^\text{prefill} &= F_{qkv}^\text{prefill}+F_{attn}^\text{prefill}+F_{feed}^\text{prefill}\\
    &= 2\left[\left(2+2\frac{g}{d}+3m\right)H+2{s_{in}}\right]s_{in}H
\end{aligned}
\end{equation}

\textbf{Decoding stage:} The reasoning process in this stage is similar to that in the prefill stage. The difference is that in the Decoding stage, each output token is generated in an autoregressive manner, and the KV cache is constantly updated. Suppose that the token embeddings input in the current Transformer layer are denoted as $t\in \mathbb{R}^{1\times H}$. First, the KV cache needs to be updated:
\begin{equation}\label{decoding_kv}
    K \leftarrow concat\left(K,t\cdot W_K\right), \quad V \leftarrow concat\left(V,t\cdot W_V\right)
\end{equation}

The subsequent calculations are consistent with the reasoning process in the Prefill stage, as follows:
\begin{equation}\label{decoding_q}
    Q=t\cdot W_Q
\end{equation}
\begin{equation}\label{decoding_A}
    A_i^h=softmax\left(\frac{Q_i^h\cdot \left(K_j^h\right)^\mathsf{T}}{\sqrt{h}}\right)\cdot V_j^h
\end{equation}
\begin{equation}\label{decoding_t_out}
    t_{out} = concat\left(A_1^h,A_2^h,\dots,A_d^h\right)\cdot W_O+t
\end{equation}
\begin{equation}\label{decoding_t_gate}
    t_{gate}=\text{SiLU}\left(t_{out}\cdot W_1\right)\otimes\left(t_{out}\cdot W_3\right)
\end{equation}
\begin{equation}\label{decoding_t_final}
    t_{final} = t_{gate}\cdot W_2+t_{out}
\end{equation}

The Decoding stage will update the KV cache and generate the remaining $s_{out}-1$ tokens in an autoregressive manner. During the autoregressive inference process, when generating the $(k+1)$-th token, the dimensions of K and V are $[s_{in}+k, h_g]$. The length of the output sequence is known to be $s_{out}$, and the prefill stage will generate the first output token. Therefore, in the Decoding stage, the dimensions of K and V will gradually increase from $[s_{in}+1, h_g]$ to $[s_{in}+s_{out}-1, h_g]$.

Based on Eq. (\ref{decoding_kv}) and (\ref{decoding_q}), the computational costs for querying, key retrieval, and value calculation during the generation of the $(k + 1)$-th token can be expressed as $F_{qkv}^{decoding}$:
\begin{equation}
    F_{qkv}^{decoding}=2H^2+4Hgh
\end{equation}

Based on Eq. (\ref{decoding_A}) and (\ref{decoding_t_out}), the computational cost of all the attention head output data during the generation of the $(k + 1)$-th token can be expressed as $F_{attn}^{decoding}$:
\begin{equation}
    F_{attn}^{decoding}=4(s_{in}+k)H+2H^2
\end{equation}

Based on Eq. (\ref{decoding_t_gate}) and (\ref{decoding_t_final}), the computational load in the feed forward network during the generation of the $(k + 1)$-th token can be expressed as $F_{feed}^{decoding}$:
\begin{equation}
    F_{feed}^{decoding}=6mH^2
\end{equation}

In summary, in the decoding stage, the inference computational cost of a single Transformer layer is expressed as:
\begin{equation}
\begin{aligned}
     F^\text{decoding} &= \sum\limits_{k=1}^{s_{out}-1} \left(F_{qkv}^\text{decoding}+F_{attn}^\text{decoding}+F_{feed}^\text{decoding}\right)\\
     &=2\left[\left(2+\frac{2g}{d}+3m\right)H+2s_{in}+s_{out}\right]\\
     &\quad \left(s_{out}-1\right)H
\end{aligned}
\end{equation}

By combining the inference processes of the prefill stage and decoding stage in an LLM, the total floating-point computation volume $F(s_{in},s_{out})$ generated during the inference process of an LLM with $L$-layer Transformers can be derived as follows:
\begin{equation}
\begin{aligned}
F(s_{in},s_{out}) 
&= \left(F^{\text{prefill}} + F^{\text{decoding}}\right)L \\
&= 2 \Bigg[
    \left(2 + \frac{2g}{d} + 3m\right)
    \left(s_{in} + s_{out} - 1\right) H  \\
&\quad + 2 s_{in}^2
    + \left(2s_{in} + s_{out}\right)
      \left(s_{out} - 1\right)
    \Bigg] H L
\end{aligned}
\end{equation}

Therefore, the processing rate $\mu_{i}^{ai}$ of the AI service $s_i^{ai}$ with input sequence length $s_{in}^i$ and output sequence length $s_{out}^i$ on a single GPU can be expressed as:
\begin{equation}
    \mu_{i}^{ai} = \frac{FPP}{F(s_{in}^i,s_{out}^i)}
\end{equation}

The storage requirements of an LLM primarily consists of the model weights and the KV cache. Therefore, the storage resource requirements $r_{s_i^{ai}}^{mem}$ for the AI service $s_i^{ai}$ with input sequence length $s_{in}^i$ and output sequence length $s_{out}^i$ can be expressed specifically as:
\begin{equation}
    r_{s_i^{ai}}^{mem} = m\_w+kv\_c
\end{equation}
where:
\begin{equation}
    m\_w = 2\left(2 + \frac{2g}{d} + 3m\right)H^2 L
\end{equation}
\begin{equation}
    kv\_c = 4\left(s_{in}^i+s_{out}^i-1\right)\frac{g}{d}HL
\end{equation}
and $m\_w$ represents the size of the model weight data, $kv\_c$ represents the size of the model KV cache data.

\subsection{Service Hybrid Deployment Models}
In edge service networks, the actual deployment of service requests uses services within the request chain as deployment units. Service instances (including AI services and microservices) are deployed on adjacent edge servers within the network and exclusively occupy certain server resources. Due to the limited resources on edge servers, service instances cannot be deployed in unlimited quantities on edge servers. We denote the number of instances $s$ deployed on server $n$ as $N_n^{s}$. Under the resource constraints of edge servers, we have the following constraints:
\begin{equation}\label{Constraints1}
    \forall n \in N, s\in {S^{MS}\cup S^{AI}},\quad \sum \limits_{s\in S} N_n^{s}r_{s}^{cpu}\le R_n^{cpu}
\end{equation}
\begin{equation}\label{Constraints2}
    \forall n \in N, s\in {S^{MS}\cup S^{AI}},\quad \sum \limits_{s\in S} N_n^{s}r_{s}^{gpu}\le R_n^{gpu}
\end{equation}
\begin{equation}\label{Constraints3}
    \forall n \in N, s\in {S^{MS}\cup S^{AI}},\quad \sum \limits_{s\in S} N_n^{s}r_{s}^{mem}\le R_n^{mem}
\end{equation}

\subsection{Service Request Routing Models}
Service requests are directed to service instances on edge servers in the form of traffic streams and await processing. The entry servers for each service request are denoted by $\{n^{sc_1}, n^{sc_2}, \dots\}$, with arrival rates $\{\lambda^{sc_1}, \lambda^{sc_2}, \dots\}$ at these entry points. Upon arrival, service requests are queued in the service instance queue on the server for processing. If the current server is unable to continue processing service requests, it will forward the request to a neighboring server capable of handling it. The request routing probability $p(sc)_{n_i, s_i}^{n, s}$ is defined as the probability that request $sc$ is forwarded from service $s$ on server $n$ to service $s_i$ on server $n_i$. This routing scheme is subject to the following constraints:
\begin{equation}\label{rout}
    \sum \limits_{n\in N} x_{n_i}^{s_i} e_{n_i}^{n} p(sc)_{n_i, s_i}^{n, s}=1
\end{equation}
where
\begin{equation}\label{occ}
    x_{n_i}^{s_i}=\left\{ 
    \begin{matrix}
        1, & N_{n_i}^{s_i}>0  \\
        0, & N_{n_i}^{s_i}=0  \\
    \end{matrix} \right.
\end{equation}
and
\begin{equation}\label{is_connected}
    e^{n}_{n_i}=\left\{ 
    \begin{matrix}
        1, & n\text{ and }n_i\text{ are physically connected or same}  \\
        0, & \text{otherwise}  \\
    \end{matrix} \right.
\end{equation}
where Eq. (\ref{rout}) ensures that service requests are fully routed without loss. Eq. (\ref{occ}) indicates whether the service instance $s$ is deployed on server $n$. Eq. (\ref{is_connected}) indicates whether the servers are directly reachable from each other or not.

Due to service reuse, traffic arriving at a service instance on a server may originate from distinct service requests. As illustrated in Fig. \ref{network}, service requests $sc_1$ and $sc_2$ are both routed to edge server $n_6$ for processing by service $s_7$. Following Burke's theorem and the linear additivity property of Poisson flows, request traffic can be aggregated at service instances. We define $\lambda_n^{s}$ as the average arrival rate of service requests at service instance $s$ on the server $n$, which is expressed as follows:
\begin{equation}\label{ms_lambda}
    \lambda_n^{s} = \sum \limits_{sc\in SC'} x_n^{s} \lambda^{sc} + \sum \limits_{sc\in SC} \sum \limits_{s\in S} \sum \limits_{n\in N}x_n^{s} p(sc)_{n, s}^{n', s'}\lambda_n^{s} 
\end{equation}
where $SC'=\{sc|ch(sc)[0]=s,sc\in SC\}$. 

Due to request routing decisions, service requests may follow multiple feasible processing paths within the network. The set of processing paths for service request $sc$ is denoted by $RP(sc)$. Each path $rp(sc)=\{n^1,n^2,\cdots\}$ in $RP(sc)$ represents a specific routing sequence, whose length corresponds to the microservice chain length in the service request, i.e., $|rp(sc)| = |sc|$. As depicted in Fig. \ref{network}, service request $sc_1$ with service chain $ch(sc_1)\{s_1,s_2,s_4,s_7\}$ has feasible processing paths $rp(sc)=\{n_5,n_2,n_2,n_1\}$ and $rp(sc)=\{n_5,n_7,n_7,n_6\}$. Let $P_{sc}^{rp(sc)}$ denote the probability that service request $sc$ completes all service processing along routing path $rp(sc)$. This probability is formally expressed as follows:
\begin{equation}
    \forall rp(sc) \in RP(sc), P(sc)_{rp(sc)} = \prod_{i=0}^{|rp(sc)|-1} p({sc})_{n^{i+1}, s_{sc}^{i+1}}^{n^{i}, s_{sc}^{i}}
\end{equation}
where $n^i$ and $n^{i+1}$ are neighboring servers in $rp(sc)$, and $s_{sc}^i$ and $s_{sc}^{i+1}$ are neighboring services in service request $sc$.
\subsection{Service Request Response Delay Analysis}
The complex process of handling service requests comprises four fundamental stages: request transmission, queuing and processing, communication forwarding, and result return. 
\subsubsection{\textbf{Transmission delay}}
The process of transmitting service requests to servers via wireless access networks includes the uploading of service request data packets and their propagation in wireless media. Given that propagation delay in wireless media is significantly shorter than the transmission duration, we approximate the transmission delay of the service request as the aggregate time spent transmitting the data packet. The delay for a user to send a service request $sc$ to the ingress server is denoted by $T^{t}_{sc}$:
\begin{equation}
    T^{t}_{sc} = \frac{D_{sc}}{B_{n,sc}}
    \label{send delay}
\end{equation}
where $D_{sc}$ denotes the packet size of service request $sc$ and $B_{n,sc}$ denotes the bandwidth between server $n$ and user $u_{sc}$ who sends service request $sc$.
\subsubsection{\textbf{Queueing and processing delay}}
The process of routing a large number of service requests into the service instance queue for processing constitutes a random service system. Thus, the queuing behavior of service requests within the service instance queue can be modeled using an Open Jackson queuing network. This queuing process introduces both queuing delay and processing delay.

Since service instances in the network encompass both AI service instances and microservice instances, the processing rate of service instances should be expressed as follows:
\begin{equation}
\mu\left(s\right)=\left\{ 
    \begin{matrix}
        \mu_i^{ms}, & s\in S^{MS}  \\
        \mu_i^{ai}, & s\in S^{AI}  \\
    \end{matrix} \right.
\end{equation}

To maintain the stability of the system, we require that the service intensity on the service instances in the servers can not exceed 1. This is mathematically expressed as:
\begin{equation}\label{rho}
    \forall s\in S^{MS}\cup S^{AI},\forall n\in N, \rho_n^{s} = \frac{\lambda_n^{s}}{N_n^{s}\mu\left(s\right)} <1
\end{equation}
where $\mu\left(s\right)$ represents the processing rate of service $s$.

We define $T_{n, s}^{q,p}$ to denote the queuing and processing latencies of a service request in the queue of service instances $ms$ on the server $n$. Therefore, according to the M/M/C queuing network \cite{M/M/C}, $T_{n, s}^{q,p}$ is obtained by the following equation:
\begin{equation}
    T_{n, s}^{q,p} = \frac{p_{v}^{s} \rho_{v}^{s}}{\lambda_{n}^{s} N_{n}^{s}!\left(1-\rho_{n}^{s}\right)^{2}}\left(\frac{\lambda_{n}^{s}}{\mu\left(s\right)}\right)^{N_{n}^{s}} +\frac{1}{\mu\left(s\right)}
\end{equation}
where 
\begin{footnotesize}
   \begin{equation}
    p_{v}^{s}=\left[\sum_{k=0}^{N_{n}^{s}-1} \frac{1}{k!}\left(\frac{\lambda_{n}^{s}}{\mu\left(s\right)}\right)^{k}+\frac{1}{N_{n}^{s}!\left(1-\rho_{n}^{s}\right)}\left(\frac{\lambda_{n}^{s}}{\mu\left(s\right)}\right)^{N_{n}^{s}}\right]^{-1}
\end{equation} 
\end{footnotesize}
\subsubsection{\textbf{Communication delay}}
When servers can process dependent services continuously within a service chain, the communication delay between them may be negligible. Otherwise, the processing results of a service instance must be transmitted to the next-hop server. We define the resultant data size of the service $s$ as $D_s$. The communication delay between server $n$ and server $n'$ is denoted by $T_{n,n'}^{c}$, which is calculated as follows:
\begin{equation}
    T_{n,n'}^{c} = \frac{D_{s}}{B_{n,n'}}
    \label{communication delay}
\end{equation}
\subsubsection{\textbf{Return delay}}
Upon completing the last service task within a service request, the server packages the processing results and returns them to the requesting user via wireless transmission. This result return process mirrors the transmission mechanism used for service request delivery. We define $D_{sc}^{return}$ as the size of the result data for service request $sc$. Consequently, the delay $T_{sc}^{r}$ incurred in returning the results of service request $sc$ is expressed as:
\begin{equation}
    T_{sc}^{r} = \frac{D_{sc}^{return}}{B_{n,sc}}
    \label{reception delay}
\end{equation}
where $B_{n,sc}$ denotes the bandwidth between server $n$ and user $u_{sc}$ who sends service request $sc$.

\subsubsection{\textbf{Average service request response delay}}
Combining the above discussion with the service request routing policy, we define the \textbf{average service response delay} $\mathbb{E}\left[T(sc) \right]$ for a service request $sc$ as follows:
\begin{equation}
\begin{aligned}
\mathbb{E}\left[T(sc) \right] 
&=\sum\limits_{rp(sc) \in RP(sc)} \Bigg[P(sc)_{rp(sc)} 
\Bigg(T_{sc}^{t}+ T_{sc}^{r} \\
&\quad+ \sum \limits_{\genfrac{}{}{0pt}{}{n\in rp(sc)}{s \in ch(sc)}} T_{n,s}^{q, p} + \sum \limits_{n,n' \in rp(sc)} T_{n,n'}^{c}\Bigg)\Bigg]
\end{aligned}
\end{equation}

\subsection{Training Objective}
The overarching objective of this paper is to minimize total service request response delay $T$ through optimizing the deployment of service instances and request routing. Based on the preceding analysis, the total service request response delay minimization problem is formulated as follows:
\begin{equation}\label{totalT}
    \mathop{\min } \sum\limits_{sc\in SC}\mathbb{E}\left[T(sc) \right]
\end{equation}
subject to (\ref{Constraints1}), (\ref{Constraints2}), (\ref{Constraints3}), (\ref{rout}), (\ref{rho}).

\section{Algorithm Design}\label{section4}
\begin{figure*}[t]
    \centering
    \includegraphics[width=182mm]{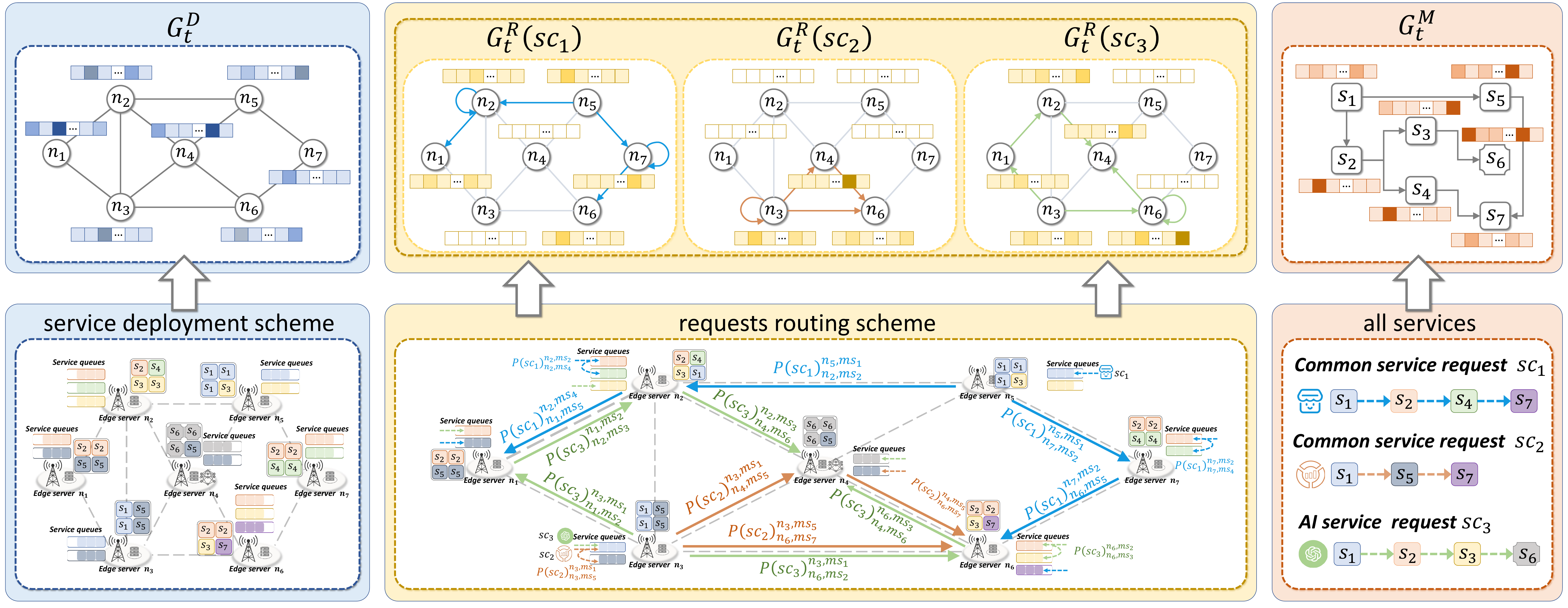}
    \vspace{-0.5cm}
    \caption{The schematic diagram of \textbf{graph-based structures $G_t^D$, $G_t^R$, and $G_t^S$}.}
    \label{Graph}
    \vspace{-0.3cm}
\end{figure*}
The close coupling between service deployment and request routing poses a significant challenge for traditional algorithms, which often struggle to optimize both aspects simultaneously and exhibit low efficiency. Reinforcement learning (RL), conversely, effectively leverages agent learning capabilities to enhance exploration efficiency and enable context-aware personalized optimization. In this context, we propose the Self-Imitation Learning-enhanced Graph Policy Optimization Algorithm (SIL-GPO) for generating static services deployment strategies and requests routing strategies.
\subsection{Interactive Environment}
The SIL-GPO algorithm enables agents to learn optimal strategies for maximizing cumulative rewards through continuous interaction with the environment. To achieve low-delay service deployment policies and request routing policies, i.e., effectively maximizing cumulative rewards, the algorithm must map states to optimal action decisions. Consequently, we model the continuous agent-environment interaction as a Markov Decision Process (MDP). At each step $t$, the agent observes the current state $s_t$, executes an action $a_t$ according to its policy $\pi(a_t|s_t)$, receives an immediate reward $r(s_t, a_t)$, and subsequently observes the resulting new state $s_{t+1}$. This interaction yields a transition $(s_t, a_t, r(s_t, a_t), s_{t+1})$ per timestep. The decision policy $\pi(a_t|s_t)$ defines a probability distribution over possible actions given the current state. The objective of the agent is to maximize the expected cumulative discounted return. Consequently, the optimal policy $\pi^*$ maximizes the expected cumulative return: \scalebox{0.8}{$\mathbb{E}_{\pi^*}\left[ \sum_{t=1}^{T} \gamma^{t} r(s_t, a_t) \right]$} where $\gamma \in [0, 1]$ denotes the discount factor. The subsequent section details the design of the three core environmental components: states, actions, and rewards.
\subsubsection{\textbf{State}}
The state perceived by the reinforcement learning agent comprises two vector-based components and three graph-based structures: The average arrival rate distribution of the service deployed in the current timestep on each server, the mark of availability for each server, the service deployment topology graph, the routing forwarding graphs, and the service invocation graph. Then, the state is represented as:
\begin{equation}\label{state}
    State = \{s_t|s_t=(\mathcal{S}_t^T,\mathcal{S}_t^R,G_t^D,G_t^R,G_t^S)\}
\end{equation}
where $\mathcal{S}_t^T$ denotes the average arrival rate distribution of the service deployed in the current timestep on each server, while $\mathcal{S}_t^R$ indicates the mark of availability for each server. $G_t^D$, $G_t^R$, and $G_t^S$ represent the service deployment topology graph, the routing forwarding graphs, and the service invocation graph, respectively. The generation processes of $G_t^D$, $G_t^R$, and $G_t^S$ are shown in Fig. \ref{Graph}. Subsequently, we will provide a detailed introduction to the various components of state.
\begin{enumerate}[label=\alph*)]
    \item \textbf{The average arrival rate distribution ($\mathcal{S}_t^T$)}: $\mathcal{S}_t^T$ denotes the current arrival rate of the service on the different servers, which is represented as:
    \begin{equation}
    \mathcal{S}_t^T = [\lambda_{n_1}^{s}, \lambda_{n_2}^{s}, \dots, \lambda_{n_{|N|}}^{s}]
    \end{equation}
    where $\lambda_{n_i}^{s}$ indicates the traffic distribution of the service instance $s$ that the agent will deploy in the current timestep.
    
    \item \textbf{The mark of availability for each server ($\mathcal{S}_t^R$)}: $\mathcal{S}_t^R$ indicates whether the remaining resources of the server can meet the service deployment requirements for this step, which is represented as:
    \begin{equation}
        \mathcal{S}_t^R = [\mathfrak{r}_1, \mathfrak{r}_2, \dots, \mathfrak{r}_{|N|}]
    \end{equation}
    If the residual resources of the server $n_i$ satisfy all resource requirements for this step of service deployment, then $\mathfrak{r}_i=1$; otherwise, $\mathfrak{r}_i=0$.

    \item \textbf{The service deployment topology graph ($G_t^D$)}: As shown in Fig. \ref{Graph}, $G_t^D=G(N_D, E_D)$ denotes the service instance deployment scheme on each server in the network, where $N_D$ denotes the set of servers in the network and $E_D$ denotes the connection between the servers. The node features $\mathcal{F}(n_d)$ of each node $n_d\in N_D$ is represented as follows: 
    \begin{equation}
        \mathcal{F}(n_d) = [x_{n_d}^{s_1}, x_{n_d}^{s_2}, \dots, x_{n_d}^{s_{|S|}}]
    \end{equation}
    Where $x_{n_d}^{s_i}$ denotes the number of services $s_i$ deployed on server $n_d$.

    \item \textbf{The routing forwarding graphs ($G_t^R$)}: As shown in Fig. \ref{Graph}, the set \scalebox{0.8}{$G_t^R=\{G_t^R(sc)|G_t^R(sc)=G(N_R, E_R)\}$} represents the route forwarding graphs for service requests $sc$ that require processing the service deployed in the current timestep, where $N_D$ denotes the set of servers in the network, and an edge $e_{ij} \in E_D$ indicates that the service request $sc$ must be forwarded from server $n_i$ to server $n_j$. Crucially, $G_t^R$ comprises multiple subgraphs. The node features $\mathcal{F}(n_r)$ of subgraph $G_t^R(sc)$ are defined as:
    \begin{equation}
        \mathcal{F}(n_r) = [p(sc)_{n_r,s_1}, p(sc)_{n_r,s_2}, \dots, p(f)_{n_r,s_{|S|}}]
    \end{equation}
    where \scalebox{0.8}{$p(f)_{n_r,s_i}=\sum_{n'\in N}p(f)_{n_r,s_i}^{n,s}$} quantifies the total probability that service request $sc$ is forwarded to service $s_i$ on server $n_r$.

    \item \textbf{The service invocation graph ($G_t^S$)}: As illustrated in Fig. \ref{Graph}, the service invocation graph $G_t^S=G(N_S, E_S)$ models dependencies among service instances, where $N_S$ represents the set of service instances and $E_S$ defines directed edges capturing invocation dependencies. The node features $\mathcal{F}(n_s)$ of each node $n_s\in N_S$ is represented as follows:
    \begin{equation}
        \mathcal{F}(n_s) = [id_{s},r_{s}^{cpu}, r_{s}^{gpu}, r_{s}^{mem},D_{s}, \mu_{s}]
    \end{equation}
    where $id_{s}$ indicates the unique identifier of service $s$; $r_{s}^{cpu}$, $r_{s}^{gpu}$, and $r_{s}^{mem}$ represent the CPU, GPU, and memory resource requirements of service $s$, respectively; $D_{s}$ denotes the output data size of service $s$; $\mu_{s}$ quantifies the processing rate of service $s$ on server.
\end{enumerate}
\subsubsection{\textbf{Action}}
The action of the agent involves deploying service instances and routing service requests to satisfy model constraints. Specifically, the agent deploys service instances across servers with heterogeneous resource capacities. To mitigate computational complexity, we design the action of the agent as an incremental deployment paradigm: each action selects exactly one service instance and deploys it on a single server within the network. Additionally, the action space employs the state variable $\mathcal{S}_t^R$ to filter invalid actions. Consequently, the action space is formulated as a discrete probability distribution:
\begin{equation}
    action = \text{softmax}\left(\mathcal{S}_t^R\odot \{a_1, a_2, \dots, a_{|N|}\}\right)
\end{equation}
where $\odot$ is the Hadamard product. 
\begin{figure*}[t]
    \centering
    \includegraphics[width=182mm]{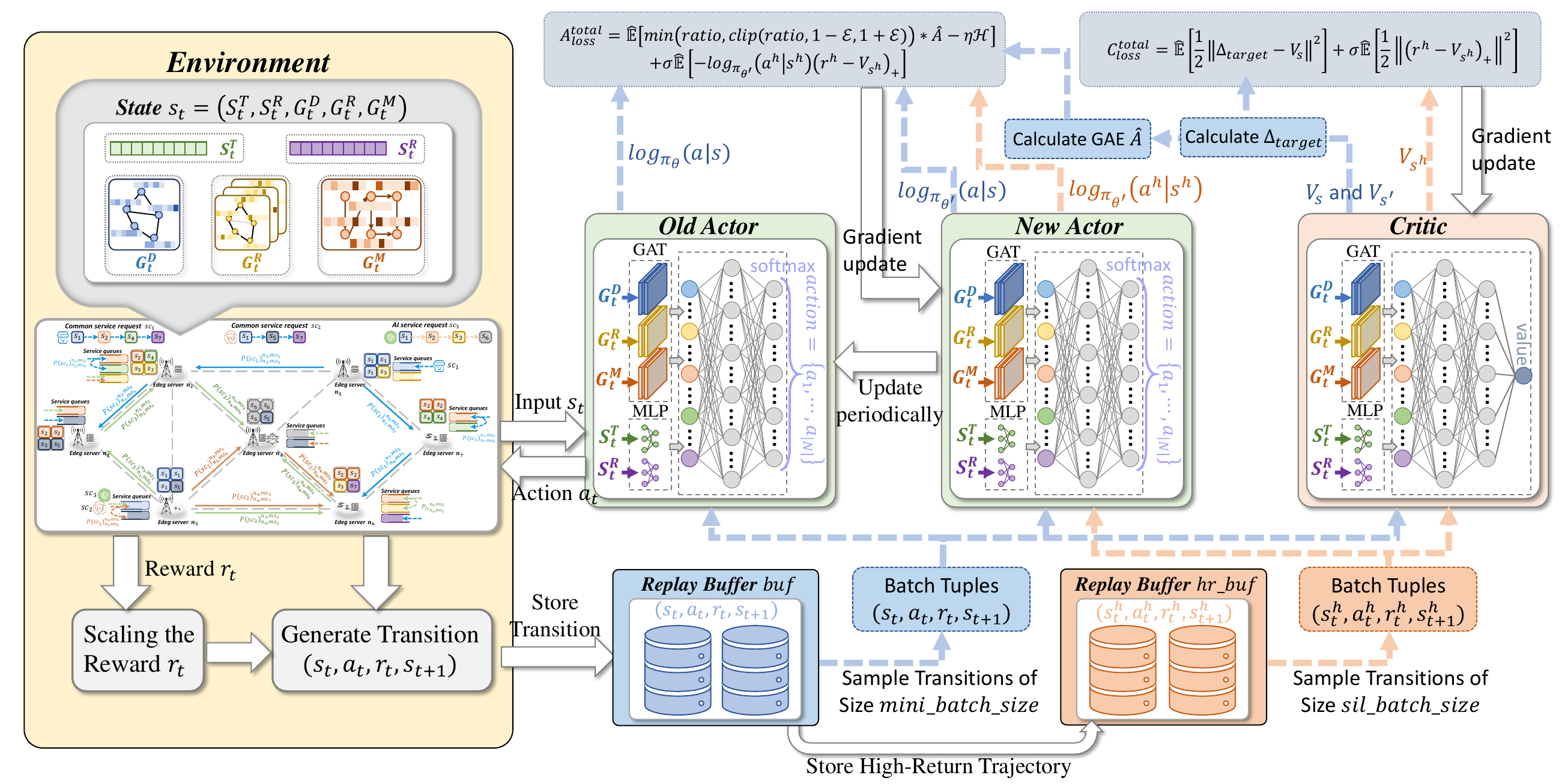}
    \vspace{-0.3cm}
    \caption{The architecture of the SIL-GPO algorithm.}
    \label{fig SIL-GPO}
    \vspace{-0.3cm}
\end{figure*}
\subsubsection{\textbf{Reward}}
The SIL-GPO algorithm, grounded in MDP, requires estimating the action-value after each agent deployment to reward or penalize the selected action. However, simultaneously deploying all service instances induces a combinatorial explosion in the action space of the agent. To mitigate this, we design the actions of the agent as incremental deployments of individual service instances. However, this sequential approach introduces a challenge: intermediate deployment actions cannot be directly evaluated, as the true end-to-end service request delay, which determines the reward signal, is measurable only after full deployment.

To address this challenge, we propose a novel dual-stage reward function comprising: Intermediate sparse rewards for each microservice deployment action, quantifying local resource utilization efficiency; Final settlement rewards based on global service delay after all deployments. This design decouples action-value estimation while preserving training convergence guarantees. Subsequently, we will introduce these two reward function designs separately.

\textbf{The intermediate sparse reward: } The intermediate sparse reward quantifies the action value during incremental microservice deployment (i.e., before all microservices are deployed). It is derived by comparing the end-to-end delay $T_i(t-1)$, which denotes the total delay after step $t-1$ in training round $i$, and $T_i(t)$, which denotes the total delay after step $t$ in training round $i$, each action:
\begin{enumerate}[label=\alph*)]
    \item A delay decrease $\left(T(t-1) \geq T(t)\right)$ yields a reward;
    
    \item A delay increase $\left(T(t-1) < T(t)\right)$ incurs a penalty;
\end{enumerate}

When the latency remains unchanged $(T(t-1)=T(t))$, it indicates that the agent deploys microservice instances on servers already hosting identical microservice types, violating the constraint in Eq. (\ref{rho}). For such actions, a fixed reward c is assigned, as horizontal scaling accelerates compliance with Eq. (\ref{rho}) and inherently enhances delay reduction potential by mitigating queuing and processing delays through resource replication. Formally, the reward function is defined as:
\begin{equation}
    R_{1} =
    \begin{cases}
        \alpha_1\left(T_i(t-1)-T_i(t)\right), & T_i(t-1)\neq T_i(t)\\
        c, & otherwise\\
    \end{cases}
\end{equation}
where $c$ is a constant reward for horizontal scaling actions and $\alpha_1$ is a positive weighting coefficient.

\textbf{Final settlement reward: }
The final settlement reward quantifies the cumulative assessment of all agent actions. Specifically, it compares the total response delay $T_{i}$ achieved in the current training round $i$ with two benchmarks: the total response delay $T_{i-1}$ from the previous round $i-1$, and the historical minimum total response delay $T_{min}$ observed during training. The reward is formulated as:
\begin{equation}
    R_2 = \alpha_2\left(T_{i-1}-T_{i}\right) + \alpha_3\left(T_{i-1}-T_{min}\right)
\end{equation}
where $T_{i-1}$ and $T_i$ represent the total response delay after completing all microservice deployments in rounds $i-1$ and $i$, respectively. $T_{min}$ denotes the minimum total response delay recorded throughout the training process.

Finally, combining these two reward calculations, our reward function can be specifically expressed as follows:
\begin{equation}\label{rewards}
    R =
    \begin{cases}
        R_1, & t<N^{total}\\
        R_1+R_2, & t=N^{total}\\
    \end{cases}
\end{equation}
where $N^{total}$ denotes the total number of service instances to be deployed.
\begin{algorithm}[htbp]
    \begin{footnotesize}
        \caption{Self-Imitation Learning-enhanced Graph Policy Optimization Algorithm (SIL-GPO).}
        \label{SIL-GPO}
        \KwIn{The cloud edge collaboration network $G(N, E)$, the set of average arrival rates for each type of microservice on each server $\lambda_n^{ms}$, discount factor $\gamma$, GAE parameter $\lambda$, clip parameter $\varepsilon$, the weight for entropy regularization $\eta$, the weight for self-imitation learning $\sigma$;}
        \KwOut{The service deployment and request routing strategy $\pi_\theta$;}
        Initialize empirical buffer $buf$, high-return empirical buffer $hr\_buf$, old actor network $\pi_{\theta}$, new actor network $\pi_{\theta'}$, critic network $V_\theta$;\\
        \For{$episode=0,1,2,\dots, max\_train\_episodes$}{
            $N^{ms} \leftarrow \lceil \frac{\sum_{n\in N}\lambda_n^{ms}}{\alpha_{ms}}\rceil$;\\
            $total\_steps = \sum_{ms \in MS}N^{ms}$;\\
            Deploy microservice instance initial solution;\\
            Initial state $s_t \leftarrow ENV.reset()$;\\
            \For{$step = 0,1,2,\dots, total\_steps$}{
                \textcolor{blue}{/* Collate the trajectory of the agent;*/}\\
                $mark(a_t) \leftarrow s_t[|N|:|N|*2]$;\\
                \If{$episode \% determining\_sample\_freq=0$}{
                    Select the action with the highest probability as the deployment action \scalebox{0.8}{$a_t\leftarrow argmax(\pi_{\theta}(a|s_t, mark(a_t)))$};\\
                }
                \Else{
                    Sample deployment actions from action probabilities $a_t\leftarrow sample(\pi_{\theta}(a|s_t,mark(a_t)))$;\\
                }
                Update the routing policy and calculate the total delay;\\
                Calculating the reward value $r_t$ by Eq. (\ref{rewards});\\
                Obtain new state $s_{t+1}$;\\
                $buf.store(s_t,a_t,r_t,s_{t+1})$;\\
                \If{$buf.count \geq batch\_size$}{
                    \textcolor{blue}{/* update parameter of the actor-critic network;*}/\\
                    Call Algorithm \ref{higt_return_buffer} to generate $hr\_buf$;\\
                    Calculate all $\Delta^{target}_t$ by Eq. (\ref{TD_error});\\
                    Calculate all $\hat{A_t}$ by Eq. (\ref{GAE});\\
                    \For{$k=0,1,2,\dots, k\_epochs$}{
                        Random sampling transitions $\mathcal{B}=\{(s_t,a_t,r_t,s_{t+1})\}$ from $buf$;\\
                        Calculate \scalebox{0.8}{$\mathcal{A}_{loss}$} and \scalebox{0.8}{$\mathcal{C}_{loss}$} by Eq. (\ref{A_loss}) and Eq. (\ref{C_loss});\\
                        Random sampling transitions $\mathcal{B}_{sil}=\{(s_t^h,a_t^h,r_kt^h,s_{t+1}^h)\}$ from $hr\_buf$;\\
                        Calculate \scalebox{0.8}{$\mathcal{A}_{loss}^{sil}$} and \scalebox{0.8}{$\mathcal{C}_{loss}^{sil}$} by Eqs. (\ref{A_sil_loss}) and (\ref{C_sil_loss});\\
                        Calculate \scalebox{0.8}{$\mathcal{A}_{loss}^{total}$} and \scalebox{0.8}{$\mathcal{C}_{loss}^{total}$} by Eqs. (\ref{A_total_loss}) and (\ref{C_total_loss});\\
                        Update the actor and critic networks using $\mathcal{A}_{loss}^{total}$ and $\mathcal{C}_{loss}^{total}$, respectively;
                    }
                    Update old actor network $\pi_\theta$ from new actor network $\pi_{\theta'}$                 
                }
            }
        }
        \textbf{return} $\pi_\theta$;
    \end{footnotesize}
\end{algorithm}
\subsection{Model Training Process}
Fig. \ref{fig SIL-GPO} illustrates the training process of the SIL-GPO algorithm for the actor and critic networks. The actor network takes the network state $s_t$ as input, generates the deployment and routing actions $a_t$ for the current state, and applies $a_t$ to the environment. Subsequently, the environment updates its state and computes the immediate reward $r_t$. After applying the sparse reward shaping function, the transition $(s_t,a_t,r_t,s_{t+1})$ is stored in the experience replay buffer $buf$. Currently, a high-return experience buffer $hr\_buf$ extracts trajectories with maximum cumulative rewards from $buf$ for prioritized storage. Finally, both the actor and critic networks sample the transitions from $buf$ and $hr\_buf$ to optimize their parameters, thereby discovering deployment and routing actions that maximize the objective function.

The specific training procedure is delineated in Algorithm \ref{SIL-GPO}. The inputs to Algorithm \ref{SIL-GPO} comprise: the edge computing network $G(N, E)$, the set of average arrival rates $\lambda_n^{ms}$for each type of microservice on each server, the discount factor $\gamma$, the GAE parameter $\lambda$, the clip parameter $\varepsilon$, the weight $\eta$ for entropy regularization, and the weight $\sigma$ for self-imitation learning. The output of Algorithm \ref{SIL-GPO} is the derived service deployment and request routing strategy $\pi_\theta$.

Algorithm \ref{SIL-GPO} commences by initializing the following components: the empirical buffer $buf$, the high-return empirical buffer $hr\_buf$, the old actor network $\pi_\theta$, the new actor network $\pi_{\theta'}$, and the critic network $V_\theta$. Subsequently, the algorithm performs multiple training iterations to jointly optimize the parameters of the actor network and the critic network and converges on the optimal microservice deployment and request routing strategy.
\begin{algorithm}[t]
    \begin{footnotesize}
        \caption{Generate a high return experience replay buffer.}
        \label{higt_return_buffer}
        \KwIn{The empirical buffer $buf$, high return ratio $\xi$;}
        \KwOut{The high-return empirical buffer $hr\_buf$;}
        $traj\_num \leftarrow \frac{batch\_size}{mini\_batch\_size}$;\\
        \For{$traj_{id} = 0,1,2,\dots, traj_{num} $}{
            According to subscript $[traj_{id}:traj_{id}+total\_steps]$, sampling trajectory $\{(s_t,a_t,r_t,s_{t+1})\}$ from $buf$;\\
            Generate the set of discounted return $[G_t,G_{t+1},\dots]$, where $G_t=\sum_{i=t}^{total\_steps}\gamma^{i-t}r_i$;\\
            Calculate the total discounted return $G_t^{total}=\sum_{i=t}^{total\_steps}G_i$;\\
            \If{$G_t^{total}>\xi*hr\_buf.max\_total\_return$}{
                $hr\_buf.max\_total\_return=max(G_t^{total}, hr\_buf.max\_total\_return)$;\\
                \For{$G_t\in G[G_t,G_{t+1},\dots]$}{
                    \If{$G_{index}>0$}{
                        $hr\_buf.store(s_t, a_t, G_t, s_{t+1})$;\\
                    }
                }
            }   
        }
        \textbf{return} $hr\_buf$;\\
    \end{footnotesize}
\end{algorithm}

The following section details the specific training procedure.
\subsubsection{Collection of learning experiences}

Based on their arrival rates, the algorithm initializes by computing the number of required microservice instances $N^{ms}$, which defines the maximum action steps of the agent (Lines 3-4, Algorithm \ref{SIL-GPO}). Subsequently, it executes the initial microservice instance deployment and initializes the environment state (Lines 5-6, Algorithm \ref{SIL-GPO}). To enhance convergence speed and algorithmic stability, an invalid action mask is applied to the agent \cite{algorithm_b1}. Then, the mask is recomputed during each training iteration (Line 8, Algorithm \ref{SIL-GPO}). Further, actions (deploying microservice instances) are sampled from the probability distribution output by the actor network. However, every $determining\_sample\_freq$ sample training iterations, the action corresponding to the maximum probability is selected instead. This strategy strikes a balance between promoting algorithmic convergence and maintaining sufficient exploration capability (Lines 9-12, Algorithm \ref{SIL-GPO}). Following action selection, the routing policy is updated, and the total service request response delay $T$ is calculated using Eq. \ref{totalT}. The routing probability $p(f)_{n, s}^{n_i, s_i}$ is defined as:
\begin{equation}
    p(f)_{n, s}^{n_i, s_i} = \frac{N_n^{s}}{\sum\limits_{n_j\in N} e_{n_i}^{n_j}N_{n_j}^{s}}
\end{equation}
where $N_n^{s}$ denotes the count of service instances $s$ deployed on server $n$, $\sum_{n_j\in N} e_{n_i}^{n_j}N_{n_j}^{ms}$ represents the total number of service instances $s$ deployed on servers adjacent to server $n$, and $N$ is the set of all servers.

The immediate reward $r_t$ associated with the action is then computed using the reward function (Eq. (\ref{rewards})), and the environment transitions to the next state $s_{t+1}$ (Lines 14-15, Algorithm \ref{SIL-GPO}). Subsequently, the resulting transition is stored in the experience replay buffer $buf$ (Line 16, Algorithm \ref{SIL-GPO}). Whenever the buffer occupancy reaches a predefined threshold, i.e., $buf.count \geq batch\_size$, the algorithm invokes Algorithm \ref{higt_return_buffer} to generate a high-return experience buffer $hr\_buf$ (Line 18, Algorithm \ref{SIL-GPO}) and prepares to optimize the actor network and the critic network parameters.
\subsubsection{Parameter update of the actor network and the critic network}
When the number of experiences in the experience replay buffer $buf$ reaches a threshold, i.e., $buf.count \geq batch\_size$, the algorithm commences model learning and parameter updates. Subsequently, the algorithm computes the Temporal Difference (TD) target, TD error, and Generalized Advantage Estimate (GAE) for state $s_t$ (lines 19, 20 in Algorithm \ref{SIL-GPO}). The computations are defined as follows:
\begin{equation}\label{TD_error}
    \Delta_{t}^{target} = r_{t}+\gamma V_\theta(s_{t+1})
\end{equation}
\begin{equation}\label{GAE}
    \hat{A_t} = \sum_{i=0}^{\infty}(\gamma\lambda)^{i}\delta_{t+i}
\end{equation}
where the state evaluations $V_\theta(s_{t+1})$ generated by the critic network when the state $s_{t+1}$ are fed as input. The term $\delta_{t+i}$ is computed as:
\begin{equation}
\begin{aligned}
        \delta_{t+i} &= \Delta_{{t+i}}^{target}-V_\theta(s_{t+i})\\
        &=r_{t+i}+\gamma V_\theta(s_{t+i+1})-V_\theta(s_{t+i})
\end{aligned}
\end{equation}

Afterward, the experience replay buffer $buf$ is partitioned into multiple mini-batches $\mathcal{B}=\{(s_t,a_t,r_t,s_{s+1})\}$ of size $|\mathcal{B}|$, over which $k\_epochs$ training iterations are performed (Line 22, Algorithm \ref{SIL-GPO}). Within each training iteration, the probability ratio is computed to quantify the divergence between the old and new policies $\pi_\theta$ and $\pi_{\theta'}$:
\begin{equation}\label{ratio}
    ratio_t =\frac{log\pi_{\theta'}(a_t|s_t)}{log\pi_\theta(a_t|s_t)}
\end{equation}

Furthermore, an entropy regularization term $\mathcal{H}$ is incorporated to incentivize exploration by the agent. This term is defined as the entropy of the action distribution under policy $\pi_{\theta'}$ at each timestep $t$.
\begin{equation}\label{entropy}
    \mathcal{H}_t^{\pi_{\theta'}} = -\sum_{a\in action}\pi_{\theta'}(a|s_t)log\pi_{\theta'}(a|s_t)
\end{equation}

Subsequently, the policy loss $\mathcal{A}_{loss}$ and value loss $\mathcal{C}_{loss}$ are then computed as follows (Line 23, Algorithm \ref{SIL-GPO}):
\begin{equation}\label{A_loss}
    \mathcal{A}_{loss} =\mathop{\hat{\mathbb{E}}}\limits_{(s_t,a_t,r_t,s_{t+1})\sim \mathcal{B}}\left[\Theta_t\hat{A}_t-\eta\mathcal{H}_t^{\pi_{\theta'}}\right]
\end{equation}
\begin{equation}\label{C_loss}
    \mathcal{C}_{loss} =\mathop{\hat{\mathbb{E}}}\limits_{(s_t,a_t,r_t,s_{t+1})\sim \mathcal{B}}\left[\frac{1}{2}\Vert\Delta^{target}_t-V_\theta(s_t) \Vert^2\right]
\end{equation}
where
\begin{equation}
    \Theta_t=min(ratio_t,clip(ratio_t,1-\varepsilon,1+\varepsilon))
\end{equation}
and $\varepsilon$ is the clip parameter, $\eta$ is the weight of entropy regularization, $\hat{A}_t$ is the GAE, and $\Delta^{target}_t$ is the TD target.

The algorithm subsequently draws samples mini-batch $\mathcal{B}_{sil}=\{(s_t^h,a_t^h,r_t^h,s_{t+1}^h)\}$ from the high-return experience buffer $hr\_buf$ for self-imitation learning and computes the corresponding loss of self-imitation learning (lines 24, 25 in Algorithm \ref{SIL-GPO}). The loss functions are defined as follows:
\begin{equation}\label{A_sil_loss}
    \mathcal{A}_{loss}^{sil} = \mathop{\hat{\mathbb{E}}}\limits_{(s_t^h,a_t^h,r_t^h,s_{t+1}^h)\sim \mathcal{B}_{sil}}\left[-log\pi_{\theta'}(a_t^h|s_t^h)\left(r_t^h-V_\theta(s_t^h)\right)_{+}\right]
\end{equation}
\begin{equation}\label{C_sil_loss}
    \mathcal{C}_{loss}^{sil} = \mathop{\hat{\mathbb{E}}}\limits_{(s_t^h,a_t^h,r_t^h,s_{t+1}^h)\sim \mathcal{B}_{sil}}\left[\frac{1}{2}\Vert\left(r_t^h-V_\theta(s_t^h)\right)_{+}\Vert^2\right]
\end{equation}
where $(\cdot)_{+} = max(\cdot,0)$ denotes the rectifier function, $\pi_{\theta'}$ is the updated actor network, $V_\theta(s_t^h)$ is the state-value estimate generated by the critic network $V_\theta$ with parameter $\theta$.

Leveraging high-return experiences, self-imitation learning enhances the efficiency of policy optimization and accelerates convergence while mitigating the risk of local optima entrapment \cite{algorithm_b2}. Finally, the total policy loss $\mathcal{A}_{loss}^{total}$ and value loss $\mathcal{C}_{loss}^{total}$ are formulated as weighted combinations of the PPO-Clip loss, the cross-entropy loss, and their respective self-imitation learning components (Line 26, Algorithm \ref{SIL-GPO}):
\begin{equation}\label{A_total_loss}
    \mathcal{A}_{loss}^{total} = \mathcal{A}_{loss} + \sigma \mathcal{A}_{loss}^{sil}
\end{equation}
\begin{equation}\label{C_total_loss}
    \mathcal{C}_{loss}^{total} = \mathcal{C}_{loss} + \sigma \mathcal{C}_{loss}^{sil}
\end{equation}
where $\sigma$ is the weighting coefficient for self-imitation learning.

The algorithm subsequently employs mini-batch stochastic gradient descent to update the parameters of the new actor network $\pi_{\theta'}$ and critic network $V_\theta$ (Line 27 in Algorithm \ref{SIL-GPO}). Following $k_epochs$ iterations of mini-batch training, the parameters of the old actor network $\pi_\theta$ are synchronized with the new actor network $\pi_{\theta'}$ (Line 28 in Algorithm \ref{SIL-GPO}). Through iterative refinement of these parameters across multiple training cycles, the algorithm progressively minimizes the combined policy loss $\mathcal{A}_{loss}^{total}$ and value loss $\mathcal{C}_{loss}^{total}$. Upon convergence, it yields the optimal microservice deployment and request routing policy $\pi_\theta$.
\section{Experimental Evaluation}\label{section5}
This section presents a comprehensive performance evaluation of the proposed method through extensive trace-driven simulations. We compare our approach against heuristic, meta-heuristic, and reinforcement learning algorithms. Moreover, all experimental settings are derived from real-world trace data collected in MEC scenarios.
\begin{figure}[t]
\centering
\includegraphics[width=88.9mm]{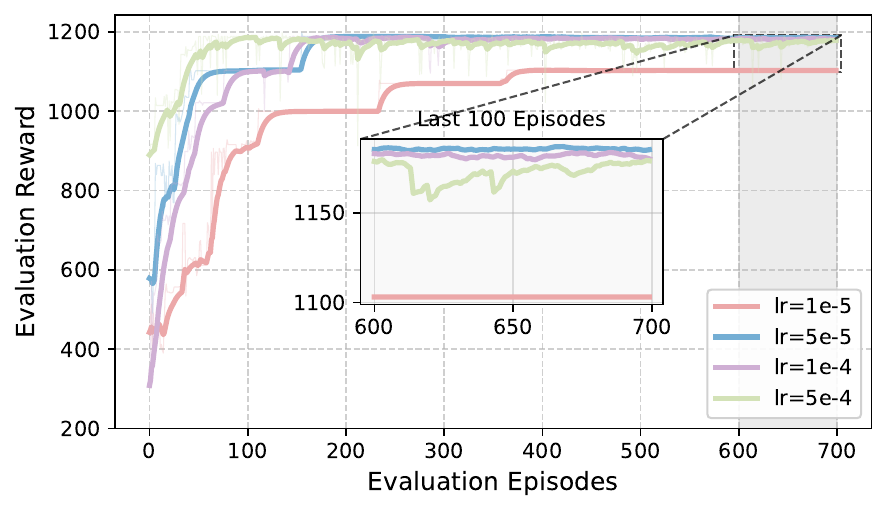}
\vspace{-0.5cm}
\caption{Reward curves under different learning rates.}
\label{fig lr}
\vspace{-0.5cm}
\end{figure}
\subsection{Experiment Settings}
All experiments were conducted on a dedicated high-performance computing server. The hardware configuration includes dual 2.50GHz, 26-core Intel(R) Xeon(R) Platinum 8269CY CPUs, dual 24GB video memory GeForce RTX 4090 GPUs, and 256GB of system memory.

\textbf{Physical networks}: 
To emulate a realistic edge computing environment, we leverage the publicly accessible EUA dataset\cite{EUA1}\cite{EUA2}, collected from real-world edge computing deployments. This dataset comprises geographic coordinates of 816 user devices and 125 edge servers within the Melbourne Central Business District (CBD). To ensure computational tractability, a representative geographical subset of this dataset was selected to instantiate our edge computing environment. Specifically, ten edge servers were deployed, consisting of seven UCS servers and three HAC servers. UCS servers are equipped with 20 CPU cores, 250 GB memory, and bandwidth capacities of 2–3 Gbps. HAC servers are configured with 30 CPU cores, 5–6 GPUs, 500 GB memory, and bandwidth capacities of 4–6 Gbps.

\textbf{Microservice instances}: Empirical studies including \cite{alibaba} indicate that deployed microservices typically span 10 to 40 instances, with 12 commonly adopted functional microservices. Furthermore, as reported in \cite{related_AI_service_deployment_b1}, contemporary applications integrate at least three categories of AI services. Based on these findings, our experimental configuration comprises 12 microservices and 3 AI services. Resource consumption and execution latency exhibit significant heterogeneity across services. In our configuration, each microservice instances are configured with 1 CPU core and 13 GB storage. Each instance generates 100 MB of data and processes requests at a service rate following a uniform distribution in [7,10]. Each AI service instances require 3 CPU cores, 1 GPU, and 65 GB storage. Each instance produces 300 MB of data and operates at a service rate uniformly distributed in [15,25].

\textbf{Service requests}: Referring to the workload characteristics and statistical patterns reported in \cite{service_steup}, we set the default number of concurrent service requests as 30, including 10 AI service requests and 20 common service requests. The service request chain length follows a uniform distribution within the range of 4 to 7. Additionally, the inter-arrival times of service requests conform to a Poisson distribution, with an average arrival rate of 2 to 4 requests per time unit.

\begin{figure}[t]
    \centering
    \includegraphics[width=88.9mm]{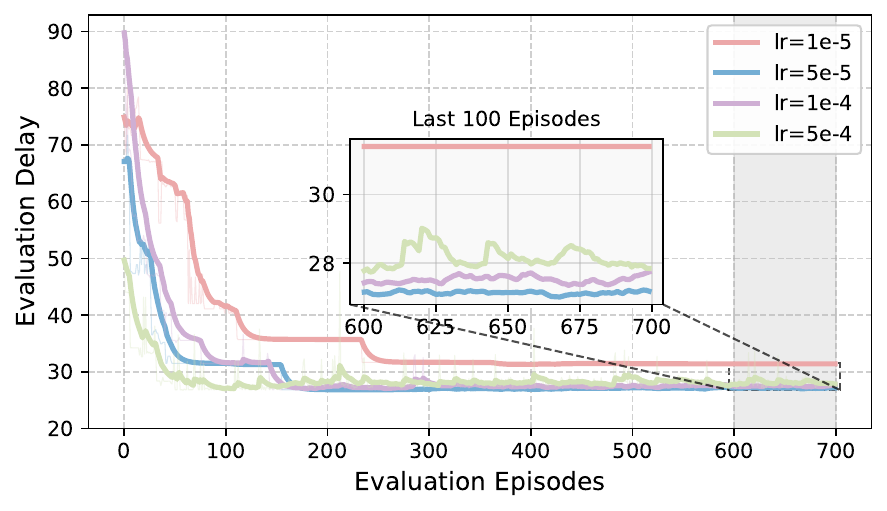}
    \vspace{-0.5cm}
\caption{Delay curves under different learning rates.}
\vspace{-0.5cm}
\label{fig de}
\end{figure}
\begin{figure}[t]
    \centering
    \subfigure[Actor loss]{
    \begin{minipage}[b]{41mm}\label{fig al}
        \centering
        \includegraphics[width=\linewidth, keepaspectratio]{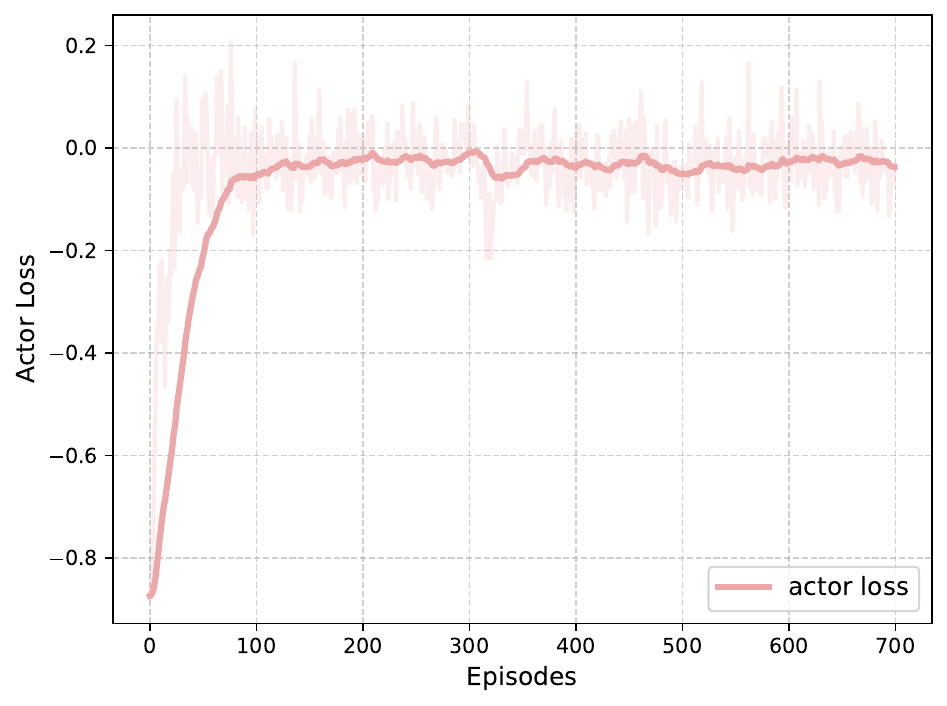}
    \end{minipage}
    }
    \subfigure[Critic loss]{
    \begin{minipage}[b]{41mm}\label{fig cl}
        \centering
        \includegraphics[width=\linewidth, keepaspectratio]{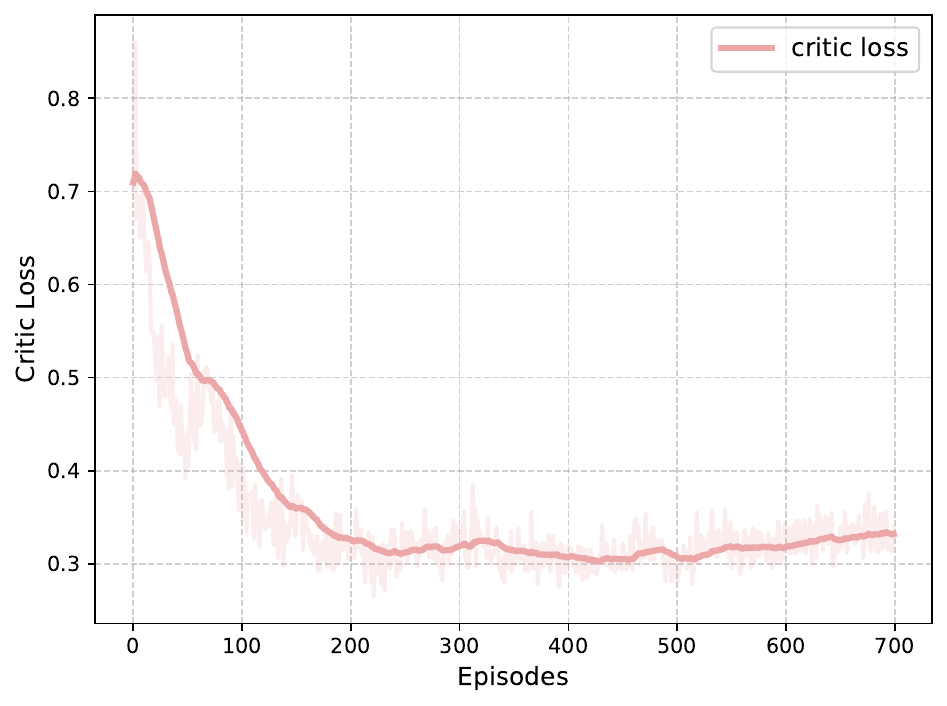}
    \end{minipage}
    }
\caption{Actor loss and critic loss of the SIL-GPO algorithm when the learning rate is 0.00005.}
\end{figure}

\textbf{Hyperparameters}: To investigate the convergence behavior of the proposed SIL-GPO algorithm, we analyze the impact of varying learning rates. As illustrated in Fig. \ref{fig lr}, increasing the learning rate from 0.00001 to 0.0001 gradually accelerates the convergence speed. However, at a learning rate of 0.00001, the converged reward value exhibits a notable decline, indicating insufficient parameter exploration that traps the model in a local optimum. As illustrated in Fig. \ref{fig lr}, empirical results demonstrate that the optimal converged reward is achieved at a learning rate of 0.00005. Furthermore, as shown in Fig. \ref{fig de}, the post-convergence delay is minimized at this learning rate. Collectively, these results establish that the optimal learning rate for convergence is 0.00005. Additionally, we present the loss trajectories of the actor and critic networks under actor and critic learning rates of 0.00005 and 0.00001, respectively. As depicted in Figs. \ref{fig al} and \ref{fig cl}, after multiple training iterations, the actor loss stabilizes near -0.05 and the critic loss converges to approximately 0.3, demonstrating the numerical stability and practical viability of SIL-GPO. The complete hyperparameter configuration is provided in Table \ref{Hyperparameter settings}.
\begin{table}[htbp]
\renewcommand{\arraystretch}{1.5}
    \centering
    \caption{Hyperparameter settings}
    \begin{tabular}{c|c}
        \hline\hline
        \textbf{Hyperparameters} & \textbf{Value}\\
        \midrule
        Learning rate for actor & 0.00005 \\
        Learning rate for critic & 0.00001 \\
        Discount factor & 0.9 \\
        GAE parameter & 0.95 \\
        PPO clip parameter & 0.2 \\
        The weight for entropy regularization & 0.05$\sim$0.1 \\
        The weight for self-imitation learning & 0.2 \\
        \hline\hline
    \end{tabular}
    \vspace{-0.3cm}
    \label{Hyperparameter settings}
\end{table}
\subsection{Baseline Algorithms}
We consider three representative baseline algorithms: a genetic algorithm-based meta-heuristic, a greedy-based heuristic, and a deep Q-learning (DQL) based on reinforcement learning. These algorithms are specifically as follows:

\textbf{HELAS} \cite{related_joint_b4}: This algorithm stochastically initializes the solution space via random search. The reciprocal of the processing delay is employed as the fitness function. A genetic algorithm is then applied iteratively to evolve the initial solution population. After each generation, the newly generated solution set is further refined using local search techniques. Through multiple iterations, the algorithm converges to a deployment strategy that minimizes the overall service processing delay.

\textbf{MFDS-FPR} \cite{MFDS-FPR}: This algorithm begins by generating an initial solution at random.  It then performs horizontal or vertical scaling of service instances based on this initial deployment. The server for scaling is selected based on the revenue of the server. By iteratively adjusting the placement of services, the algorithm identifies a deployment configuration that minimizes total response delay.

\textbf{RSDQL} \cite{related_microservice_deployment_b1}: This algorithm adopts a deep Q-learning framework that incorporates reward sharing to facilitate policy optimization. The resulting deployment policy is derived through an elastic scaling mechanism that dynamically adjusts the number of service instances. However, this approach does not support multi-instance routing. To ensure fair comparison, we adopt the routing strategy proposed in this paper.
\subsection{Experimental Results}
\begin{figure}[t]
    \centering
    \subfigure[The total delay]{
    \begin{minipage}[b]{41mm}\label{diff_arrival_rate_a}
        \centering
        \includegraphics[width=\linewidth, keepaspectratio]{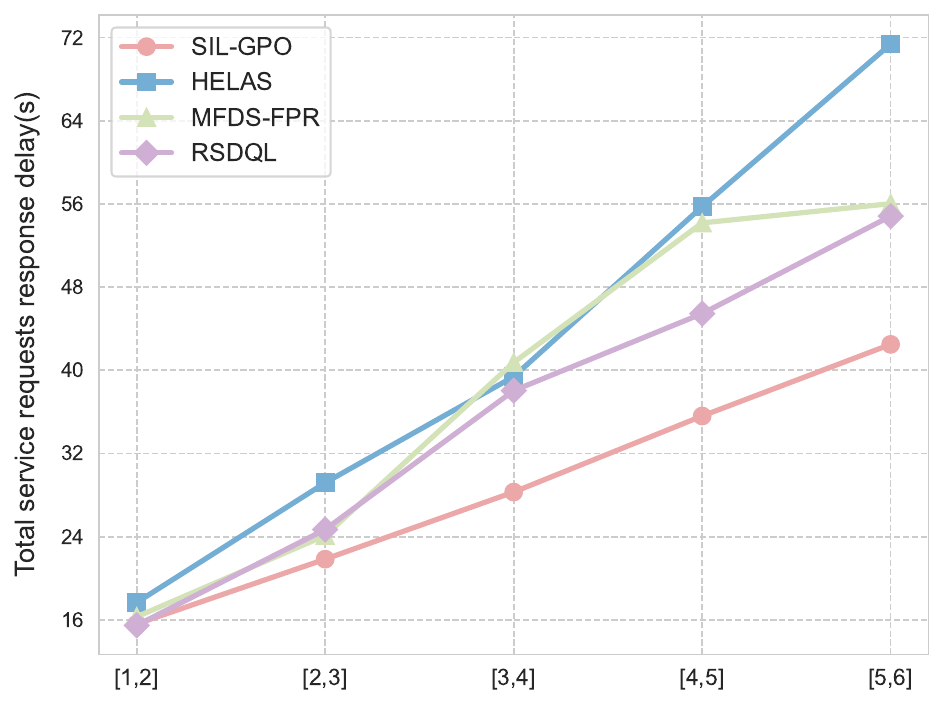}
    \end{minipage}
    } 
    \subfigure[The number of used CPU]{
    \begin{minipage}[b]{41mm}\label{diff_arrival_rate_b}
        \centering
        \includegraphics[width=\linewidth, keepaspectratio]{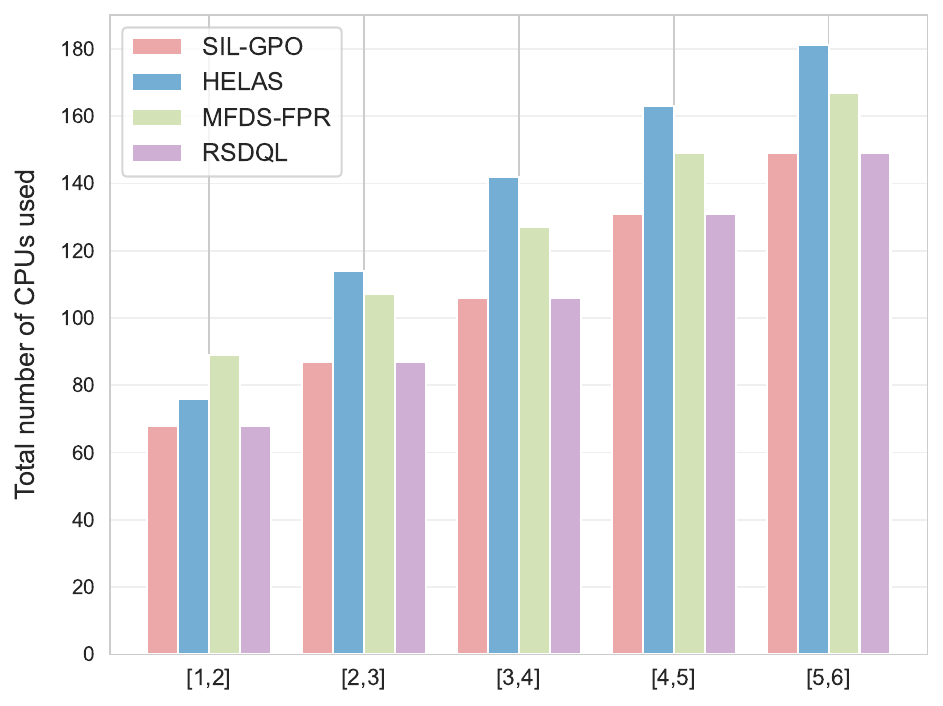}
    \end{minipage}
    }
    \\
    \subfigure[The number of used GPU]{
    \begin{minipage}[b]{41mm}\label{diff_arrival_rate_c}
        \centering
        \includegraphics[width=\linewidth, keepaspectratio]{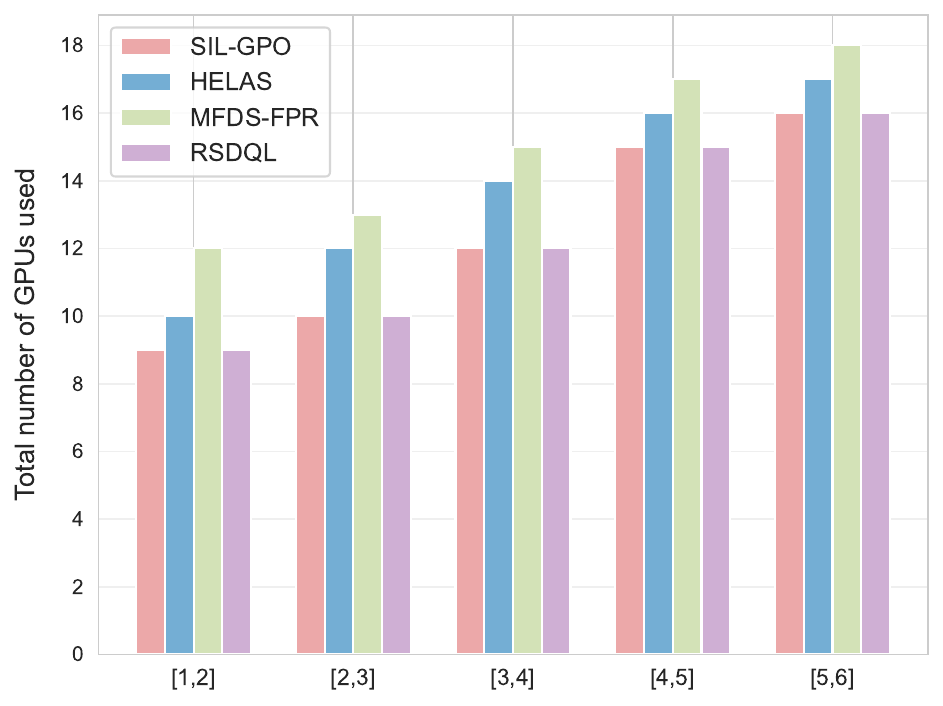}
    \end{minipage}
    } 
    \subfigure[The amount of used memory]{
    \begin{minipage}[b]{41mm}\label{diff_arrival_rate_d}
        \centering
        \includegraphics[width=\linewidth, keepaspectratio]{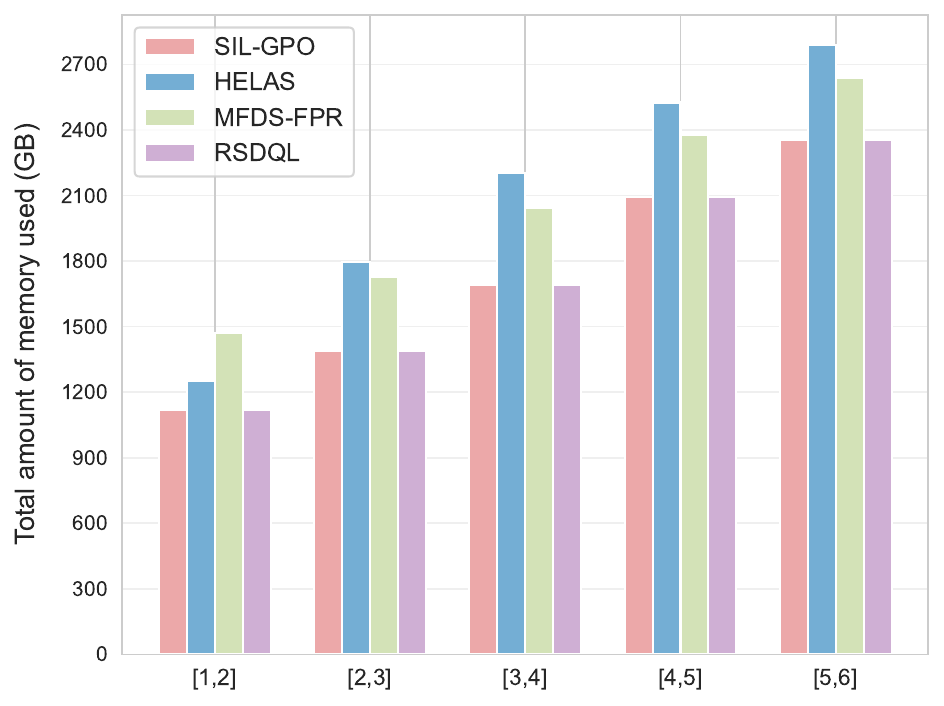}
    \end{minipage}
    }
\caption{Comparison of algorithm performance under different service request arrival rates}
\label{diff_arrival_rate}
\vspace{-0.5cm}
\end{figure}
\subsubsection{The Impact of Average Arrival Rate of Service Requests}
First, we investigate the impact of the average arrival rate of service requests on the total response delay of service requests and resource utilization. We set the average arrival rate of service requests respectively as $[1, 2]$, $[2, 3]$, $[3, 4]$, $[4, 5]$, and $[5, 6]$. As shown in Fig. \ref{diff_arrival_rate}, service response delay and resource occupancy increase as the service request arrival rate increases. As shown in Fig. \ref{diff_arrival_rate_a}, when the service request arrival rate is $[1, 2]$, the SIL-GPO algorithm and other baseline algorithms show little difference in service response delay. However, as the service request arrival rate continues to increase, the SIL-GPO algorithm demonstrates the best performance in terms of service response delay. Compared to other baseline algorithms, our algorithm achieves delay improvements of 32.6\% (HELAS), 24.8\% (MFDS-FPR), and 19.4\% (RSDQL). This is mainly because the SIL-GPO algorithm can greatly reduce waiting delays and communication delays through the perception of graph neural networks, and has better performance in reducing service request response delays. Furthermore, from Figs \ref{diff_arrival_rate_b}, \ref{diff_arrival_rate_c}, and \ref{diff_arrival_rate_d}, we can see that our algorithm and the RSDQL algorithm consume the same amount of resources, but both are lower than the HELAS algorithm and the MFDS-FPR algorithm. Although our algorithm consumes the same amount of resources as the RSDQL algorithm, it achieves lower request delay. Compared with the other two baseline algorithms, our algorithm demonstrates better performance in terms of request delay and resource consumption.
\subsubsection{The Impact of Length of Service Request Chain}
\begin{figure}[t]
    \centering
    \subfigure[The total delay]{
    \begin{minipage}[b]{41mm}\label{length_a}
        \centering
        \includegraphics[width=\linewidth, keepaspectratio]{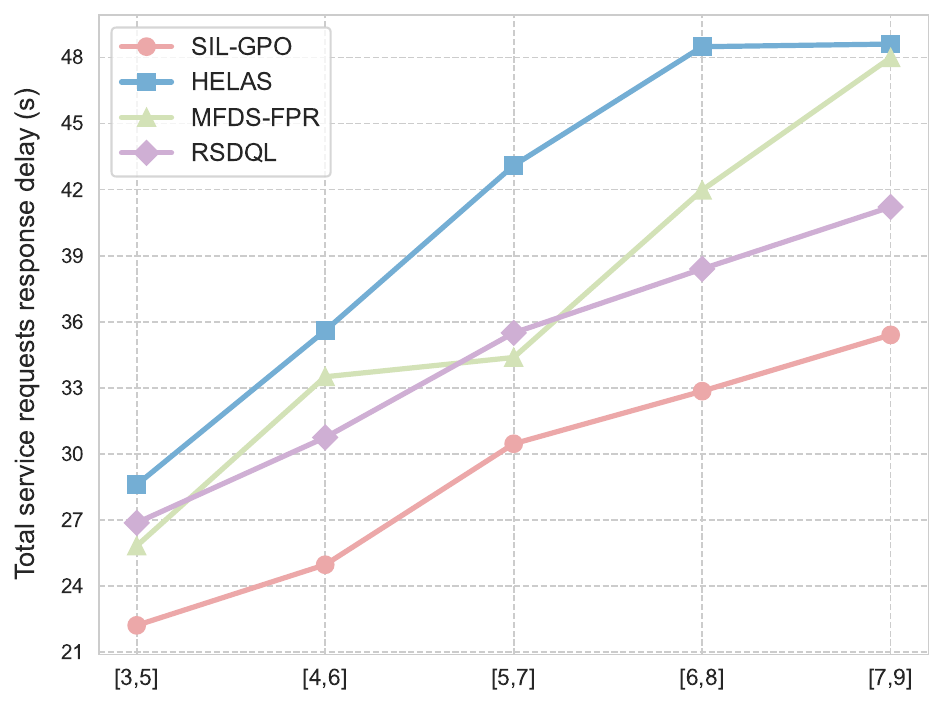}
    \end{minipage}
    } 
    \subfigure[The number of used CPU]{
    \begin{minipage}[b]{41mm}\label{length_b}
        \centering
        \includegraphics[width=\linewidth, keepaspectratio]{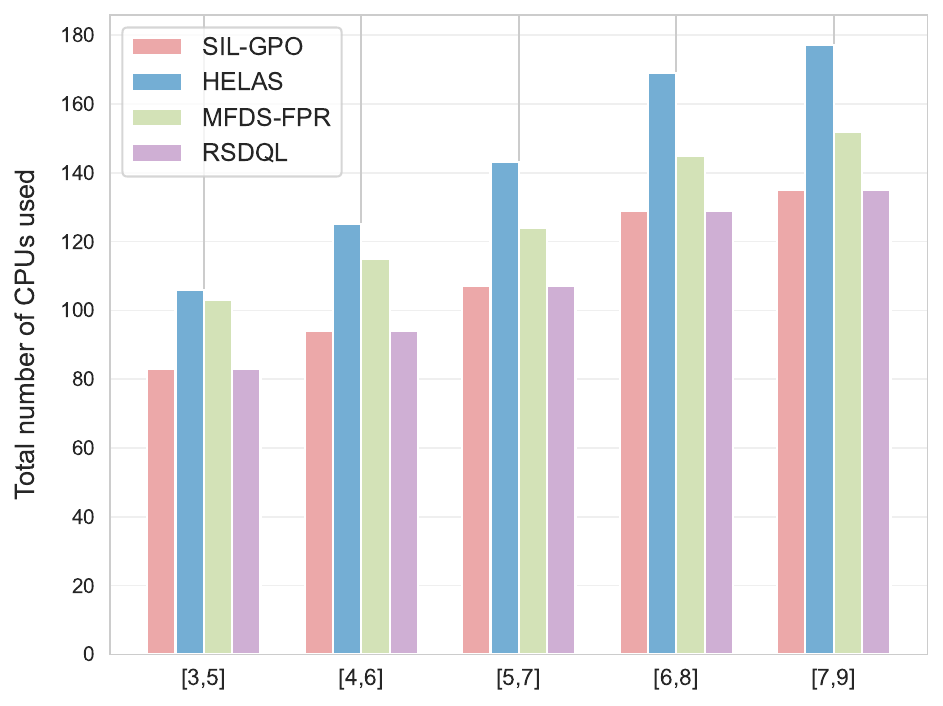}
    \end{minipage}
    }
    \\
    \subfigure[The number of used GPU]{
    \begin{minipage}[b]{41mm}\label{length_c}
        \centering
        \includegraphics[width=\linewidth, keepaspectratio]{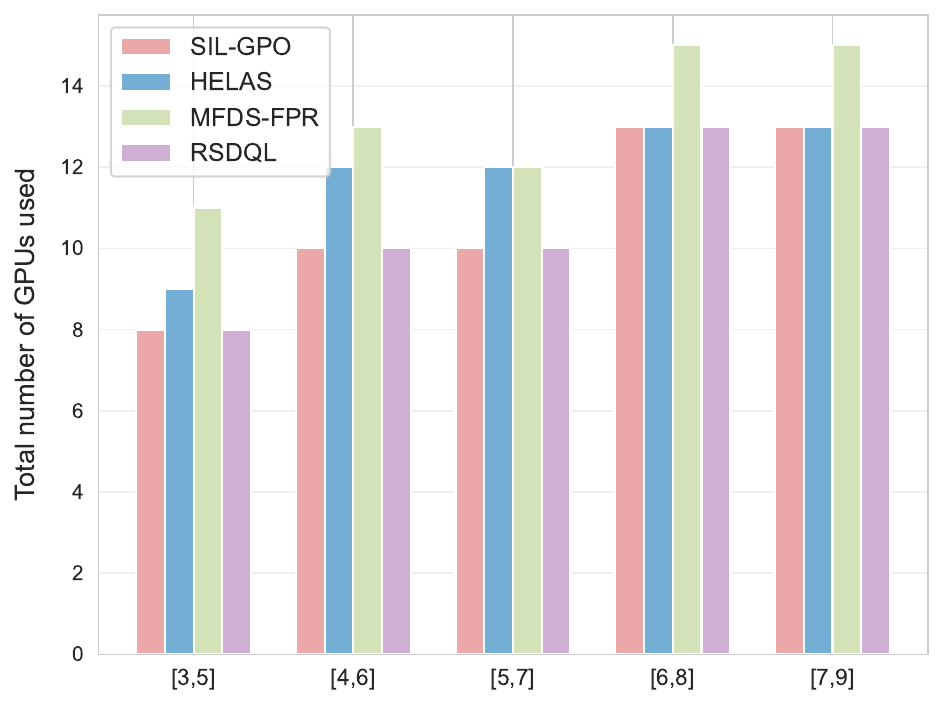}
    \end{minipage}
    } 
    \subfigure[The amount of used memory]{
    \begin{minipage}[b]{41mm}\label{length_d}
        \centering
        \includegraphics[width=\linewidth, keepaspectratio]{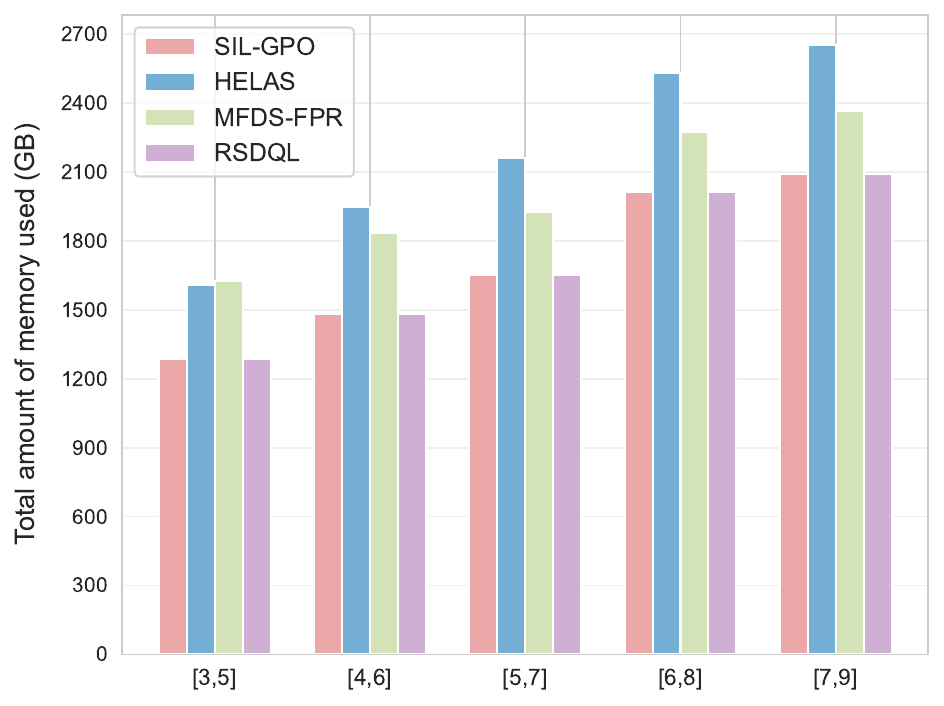}
    \end{minipage}
    }
\caption{Comparison of algorithm performance under different service request chain lengths}
\label{length}
\end{figure}

Subsequently, we analyzed the impact of the length of service requests on the total response delay and resource usage of service requests. We set the length of service requests respectively as $[3, 5]$, $[4, 4]$, $[5, 7]$, and $[6, 8]$, $[7, 9]$. As shown in Fig. \ref{length}, when the request chain length increases from $[3,5]$ to $[7,9]$, our algorithm consistently achieves the lowest request delay and the lowest usage of computing and storage resources. This is because, through feature extraction via graph neural networks, the model can always accurately assess the state of the agent, thereby guiding the agent to make the correct decision. Compared with other baseline algorithms, our algorithm achieves delay improvements of 28.6\% (HELAS), 20.7\% (MFDS-FPR), and 15.5\% (RSDQL), as shown in Fig. \ref{length_a}. Moreover, as shown in Figs. \ref{length_b}, \ref{length_c}, and \ref{length_d}, although our algorithm uses the same amount of computing and storage resources as the RSDQL algorithm, it achieves a lower delay. However, compared with the HELAS algorithm and MFDS-FPR algorithm, our algorithm not only uses fewer computing and storage resources but also achieves the lowest request delay. Since our algorithm utilizes graph neural networks to evaluate the state value of agents, it can guide agents to make correct decisions, thereby achieving service deployment schemes with lower delay and fewer resource requirements.
\subsubsection{The Impact of the Number of Service Requests}
\begin{figure}[t]\label{number}
    \centering
    \subfigure[The total delay]{
    \begin{minipage}[b]{41mm}\label{number_a}
        \centering
        \includegraphics[width=\linewidth, keepaspectratio]{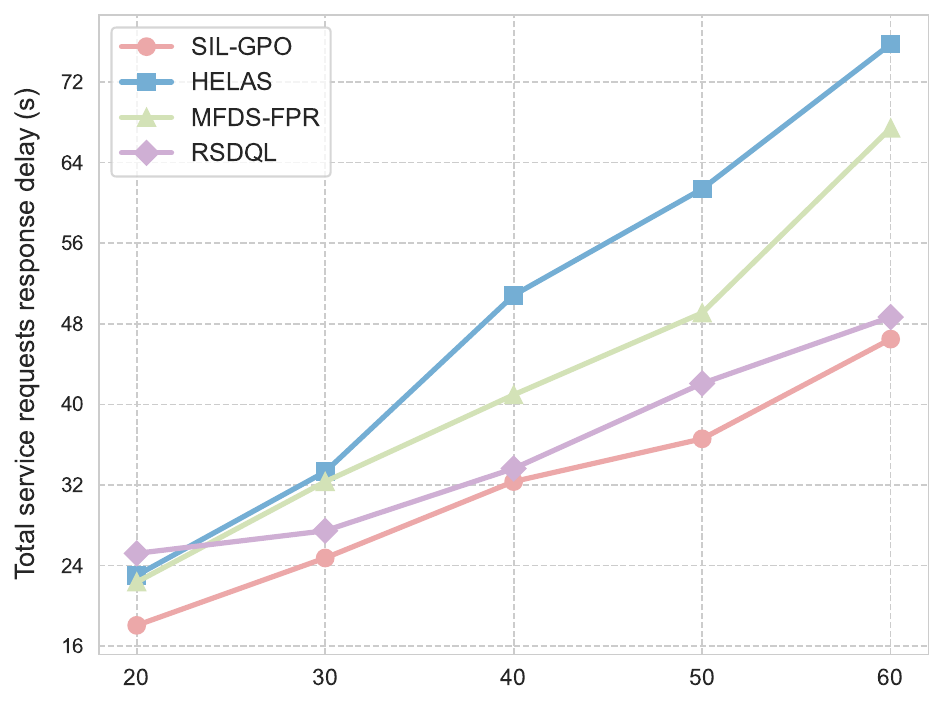}
    \end{minipage}
    } 
    \subfigure[The number of used CPU]{
    \begin{minipage}[b]{41mm}\label{number_b}
        \centering
        \includegraphics[width=\linewidth, keepaspectratio]{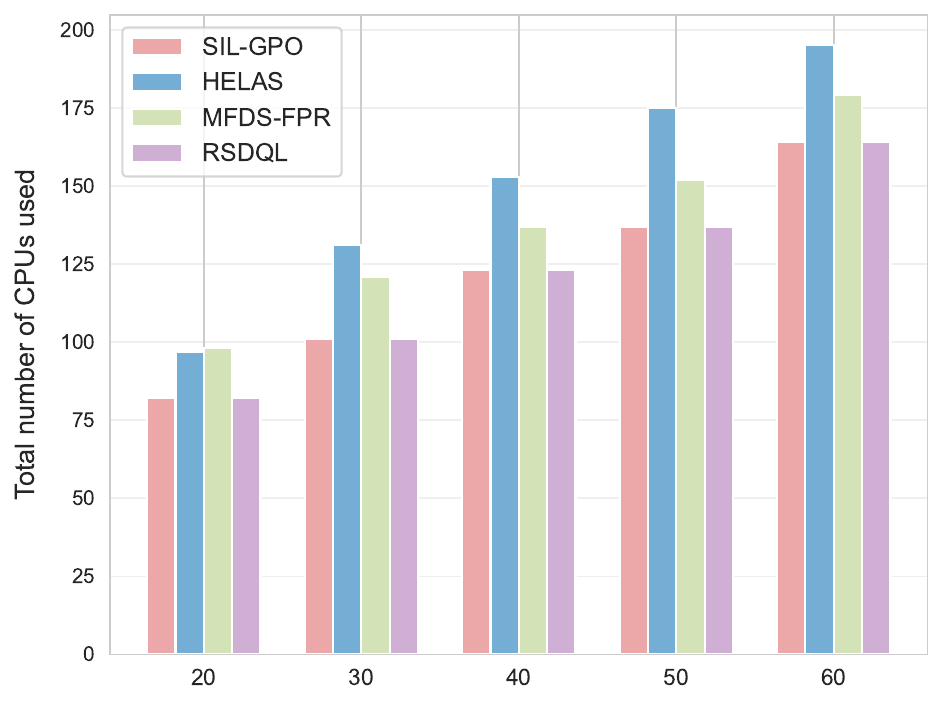}
    \end{minipage}
    }
    \\
    \subfigure[The number of used GPU]{
    \begin{minipage}[b]{41mm}\label{number_c}
        \centering
        \includegraphics[width=\linewidth, keepaspectratio]{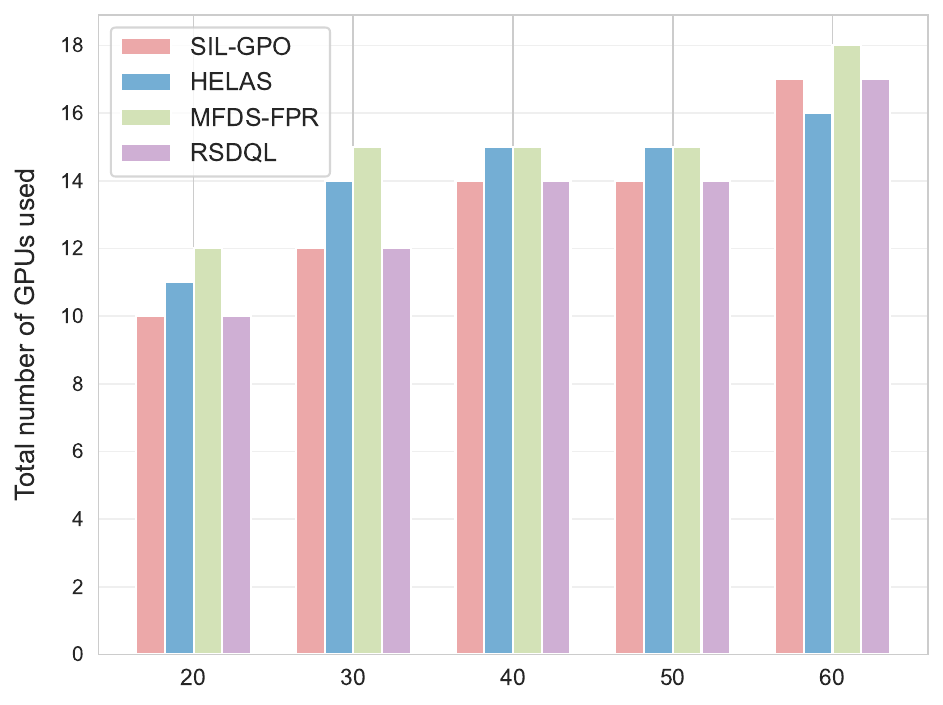}
    \end{minipage}
    } 
    \subfigure[The amount of used memory]{
    \begin{minipage}[b]{41mm}\label{number_d}
        \centering
        \includegraphics[width=\linewidth, keepaspectratio]{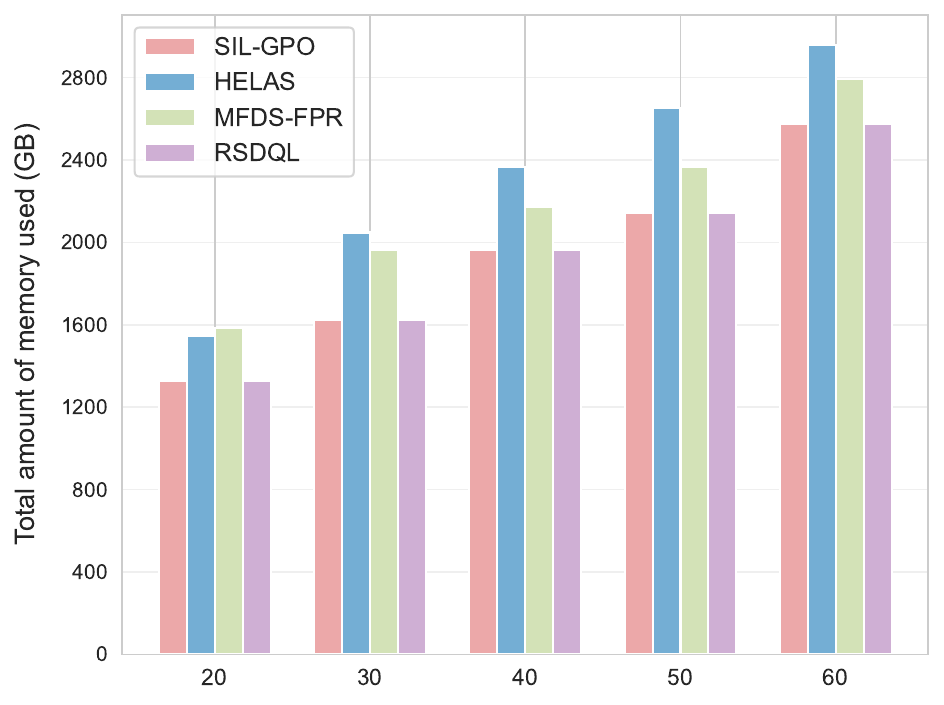}
    \end{minipage}
    }
\caption{Comparison of algorithm performance under different numbers of service request chains}
\end{figure}

Finally, we varied the number of service requests to study the impact of different numbers of service requests on the total response delay of service requests and resource usage. We gradually increased the number of service requests from 20 to 60, with a step size of 10. From Fig. \ref{number_a}, we can see that request delay increases as the number of service requests increases. However, our algorithm always achieves the lowest delay. Specifically, compared with other baseline algorithms, our algorithm reduced the total service request delay by 35.2\% (HELAS), 25.4\% (MFDS-FPR), and 10.6\% (RSDQL). In addition, our algorithm always uses the least amount of computing and storage resources from Figs. \ref{number_b}, \ref{number_c}, and \ref{number_d}. Although our algorithm and the RSDQL algorithm consume the same amount of computing and storage resources, our algorithm achieves lower delay. Compared with the other two baseline algorithms, our algorithm uses the least amount of CPU, GPU, and storage resources. This also shows that our algorithm can utilize fewer computing and storage resources and perform best in terms of service request delay.
\section{Conclusions}\label{section6}
We propose a Self-Imitation Learning-enhanced Graph Policy Optimization (SIL-GPO) algorithm to address the joint optimization of hybrid orchestration and request routing for AI services and microservices. First, the request processing workflow is modeled through an Open Jackson Network modeling framework, which formalizes four sequential stages: request transmission, queueing and processing, inter-service communication, and result return. Second, SIL-GPO integrates graph attention networks to extract agent state features and employs self-imitation learning to accelerate policy convergence. This enables the derivation of latency-optimal orchestration strategies for both service placement and request routing. Finally, SIL-GPO is rigorously evaluated against heuristic algorithms, meta-heuristic approaches, and conventional reinforcement learning methods across heterogeneous scenarios. Experimental results demonstrate that SIL-GPO consistently achieves the lowest-latency orchestration with minimal resource overhead while maintaining computational and storage efficiency.


\end{document}